\documentclass[twocolumn,aps,pra,superscriptaddress,,nofootinbib,notitlepage,showpacs,floatfix,longbibliography]{revtex4-1}

\usepackage{amsmath,amssymb,amsthm,dsfont}
\usepackage{braket}
\usepackage{empheq}
\usepackage{subfigure}
\usepackage{cancel}
\usepackage[pdfstartview=FitH]{hyperref}
\hypersetup{
    colorlinks=true,       
    linkcolor=blue,          
    citecolor=magenta,        
    filecolor=magenta,      
    urlcolor=blue,           
    runcolor= blue
}

\usepackage{cancel}
\usepackage{relsize}

\newcommand{\bes} {\begin{subequations}}
\newcommand{\ees} {\end{subequations}}
\newcommand{\beq}{\begin{equation}}
\newcommand{\eeq}{\end{equation}}
\newcommand{\ba}{\begin{eqnarray}}
\newcommand{\ea}{\end{eqnarray}}

\newcommand\norm[1]{\left\lVert#1\right\rVert}
\newcommand\eff{\mathrm{eff}}

\newcommand{\ketbra}[1]{|{#1}\rangle\langle#1|}
\newcommand{\vertiii}[1]{{\| #1\|}}
\newcommand{\ignore}[1]{}

\newtheorem*{theorem*}{Theorem}

\def\b{\beta}

\newcommand{\eps}{\varepsilon}

\begin{document}

\title{Quantum trajectories for time-dependent adiabatic master equations}

\author{Ka Wa Yip}
\affiliation{Department of Physics and Astronomy, University of Southern California, Los Angeles, California 90089, USA}
\affiliation{Center for Quantum Information Science \& Technology, University of Southern California, Los Angeles, California 90089, USA}
\author{Tameem Albash}
\affiliation{Department of Physics and Astronomy, University of Southern California, Los Angeles, California 90089, USA}
\affiliation{Center for Quantum Information Science \& Technology, University of Southern California, Los Angeles, California 90089, USA}
\affiliation{Information Sciences Institute, University of Southern California, Marina del Rey, California 90292, USA}
\author{Daniel A. Lidar}
\affiliation{Department of Physics and Astronomy, University of Southern California, Los Angeles, California 90089, USA}
\affiliation{Center for Quantum Information Science \& Technology, University of Southern California, Los Angeles, California 90089, USA}
\affiliation{Department of Electrical Engineering, University of Southern California, Los Angeles, California 90089, USA}
\affiliation{Department of Chemistry, University of Southern California, Los Angeles, California 90089, USA}


\begin{abstract}
We describe a quantum trajectories technique for the unraveling of the quantum adiabatic master equation in Lindblad form.  By evolving a complex state vector of dimension $N$ instead of a complex density matrix of dimension $N^2$, simulations of larger system sizes become feasible.  The cost of running many trajectories, which is required to recover the master equation evolution, can be minimized by running the trajectories in parallel, making this method suitable for high performance computing clusters. In general, the trajectories method can provide up to a factor $N$ advantage over directly solving the master equation. In special cases where only the expectation values of certain observables are desired, an advantage of up to a factor $N^2$ is possible. We test the method by demonstrating agreement with direct solution of the quantum adiabatic master equation for $8$-qubit quantum annealing examples. We also apply the quantum trajectories method to a $16$-qubit example originally introduced to demonstrate the role of tunneling in quantum annealing, which is significantly more time consuming to solve directly using the master equation. The quantum trajectories method provides insight into individual quantum jump trajectories and their statistics, thus shedding light on open system quantum adiabatic evolution beyond the master equation.
\end{abstract}

\maketitle

\section{Introduction}
With the growing ability to control and measure ever-larger quantum systems, understanding how to model the interactions between open quantum systems and their environment has become exceedingly important \cite{Wiseman:book}. 
The open system dynamics is often described in terms of a master equation in Lindblad form, describing the effective dynamics of the quantum system after the environmental degrees of freedom have been traced out \cite{Breuer:2002}.  An equivalent approach is that of quantum trajectories (also known as the Monte Carlo wavefunction method)~\cite{carmichael2009statistical,dum1992monte,molmer1993monte}, which can be understood as an unraveling of the master equation in Lindblad form, and which generates a stochastic process whose average is fully equivalent to the master equation (for a review, see Ref.~\cite{daley2014quantum}).  Each trajectory in this approach can also be viewed as the result of continuous indirect measurements of the environment in a certain basis \cite{brun:719}. A quantum trajectories approach exists also for non-Markovian master equations \cite{Imamoglu:94,Breuer:2004wq}. 

While a vast literature exists on the topic of quantum trajectories for time-independent master equations, much less is known for the case of time-dependent master equations (see, e.g., Ref.~\cite{Caiaffa:2017aa}), which is our focus here. Specifically, we focus on the case of open systems evolving adiabatically according to a time-dependent Hamiltonian, weakly coupled to the environment \cite{springerlink:10.1007/BF01011696,ABLZ:12-SI}. This is particularly relevant in the context of quantum annealing and more generally adiabatic quantum computing, whereby the computation proceeds via a time-dependent Hamiltonian and the result of the computation is encoded in the ground state of the final Hamiltonian (for reviews see Refs.~\cite{RevModPhys.80.1061,Albash-Lidar:RMP}). A large body of literature exists on the use of quantum Monte Carlo methods in the context of quantum annealing (see, e.g., Refs.~\cite{Santoro,Heim:2014jf,Albash:2017aa}), but these methods focus on equilibrium properties, while here we are interested in dynamics. The interplay between the key quantities that determine adiabaticity and non-unitary dynamics has not been previously explored in the setting of Monte Carlo wavefunction methods, and here we resolve this question by finding an upper bound on the size of the Monte Carlo time-step. Nor has the question of reducing the computational cost of simulations of the adiabatic master equation via quantum trajectories been discussed so far, and we address this here.

Thus, here we develop the first treatment of a quantum trajectories unravelling of a time-dependent adiabatic master equation (AME).
We make a formal comparison between the quantum trajectory unraveling of the Lindblad master equation with time-independent and time-dependent operators, and discuss the validity of applying it to the unraveling of the AME.   While our analysis closely follows the standard time-independent approach, the time-dependent case results in new additional validity conditions that must be satisfied.

The individual trajectories of the AME shed new light on how the average case captured by the AME emerges.  When the quantum state is a pure energy eigenstate and the unitary evolution is adiabatic, the drift term in the quantum trajectories approach vanishes, and the quantum state follows the instantaneous eigenstates until a jump occurs.  We can associate these jumps with an excitation or relaxation process, depending on the direction of the jump.  This provides an intuitive (yet rigorous) picture for how the averaged dynamics of the AME arises.

An important advantage of the quantum trajectories approach is that for an $N$-dimensional system, one quantum trajectory requires storing and updating $2N-1$ real numbers, while solving the master equation for a density matrix requires storing and updating $N^2-1$ real numbers. This quadratic saving allows simulations of systems with sizes that are infeasible by directly solving the master equation. The tradeoff is that many trajectories must be run in order to accurately approximate the solution of the master equation, but this tradeoff can be reduced by using many parallel processes to represent each trajectory.

Our presentation is organized as follows. In Sec.~\ref{sec:II} we briefly review the AME. We unravel the AME in Sec.~\ref{sec:III} into quantum trajectories taking the form of quantum jumps, allowing for an arbitrary time-dependence of the Hamiltonian and Lindblad operators. In Sec.~\ref{sec:IV} we provide an algorithmic implementation for our adiabatic quantum trajectories and in Sec.~\ref{sec:V} we present three case studies. We perform a cost comparison between the direct simulation of the AME and the quantum trajectories method in Sec.~\ref{sec:timequantitative}. Additional technical details and proofs are provided in the Appendices.

%

\section{Adiabatic master equation in Lindblad form}
\label{sec:II}
We focus on the AME in Lindblad form, which can be derived with suitable approximations (in the weak coupling limit after performing the Born-Markov, rotating wave, and adiabatic  approximation) from first principles starting from the system Hamiltonian $H_{\textrm{S}}$, the environment Hamiltonian $H_B$, and the interaction Hamiltonian $H_I = g \sum_\alpha A_{\alpha} \otimes B_{\alpha}$, with system operators $A_\alpha$, environment operators $B_{\alpha}$, and system-bath coupling strength $g$~\cite{ABLZ:12-SI}.  The adiabatic (Lindblad) master equation describes the evolution of the system density matrix $\rho(t)$ and has the following form (setting $\hbar = 1$ from now on):
\beq
\label{eqt:ME2-H}
\frac{d}{dt} \rho(t) = -i \left[H_{\textrm{S}}(t) + H_{\textrm{LS}}(t), \rho(t) \right] + \mathcal{L}_{\textrm{WCL}} [\rho(t)] \ , \\
\eeq
where $H_{\textrm{LS}}(t)$, which commutes with $H_{\textrm{S}}(t)$, is a Lamb shift Hamiltonian arising from the interaction with the environment.  The dissipative term  $\mathcal{L}_{\mathrm{WCL}}$ takes the form:
\begin{eqnarray}
\mathcal{L}_{\textrm{WCL}}  [\rho(t)]   &\equiv& \sum_{\alpha, \beta} \sum_{\omega} \gamma_{\alpha \beta}(\omega) \left(L_{\beta,\omega}(t) \rho(t) L_{\alpha, \omega}^{\dagger}(t)\phantom{\frac{1}{2}} \right. \notag \\
&& \hspace{-0.5cm} \left. \qquad\qquad - \frac{1}{2} \left\{ L_{\alpha,\omega}^{\dagger}(t) L_{\beta,\omega}(t) , \rho(t) \right\} \right) \ , \label{eqt:ME2-L}
\end{eqnarray}
where the sum over $\omega$ is over the Bohr frequencies (eigenenergy differences) of $H_{S}$, $\gamma_{\alpha \beta}(\omega)$ is an element of the positive matrix $\gamma$, and satisfies the Kubo-Martin-Schwinger (KMS) condition if the bath is in a thermal state with inverse temperature $\beta = 1/T$:
\beq
\gamma_{\alpha \beta}(- \omega) = e^{-\beta\omega} \gamma_{\beta\alpha }(\omega)\ .
\label{eq:KMS}
\eeq
The time-dependent Lindblad operators are given by:
\begin{align}
L_{\alpha, \omega}(t) &= \sum_{a,b} \delta_{\omega, \eps_b(t) - \eps_a(t)} \bra{\eps_a(t)} A_\alpha \ket{\eps_b(t)} | \eps_a(t) \rangle  \langle \eps_b(t)| \ ,
\label{eq:Lindblad2}
\end{align}
where $\ket{\varepsilon_a(t)}$ is the $a$-th instantaneous energy eigenstate of $H_{\textrm{S}}(t)$ with eigenvalue $\varepsilon_a(t)$. 
With this form for the Lindblad operators, decoherence can be understood as occurring in the instantaneous energy eigenbasis~\cite{Albash:2015nx}. 

For the purpose of unravelling the above master equation into quantum trajectories, it is convenient to diagonalize the matrix $\gamma$ 
by an appropriate unitary transformation $u(\omega)$:
\begin{equation}
\sum_{\alpha, \beta} u_{i, \alpha}(\omega) \gamma_{\alpha \beta}(\omega) u_{j, \beta}(\omega)^{\dagger} = \left(\begin{matrix} \gamma'_1(\omega) & 0 &\hdots \\0 & \gamma'_2(\omega) & \hdots \\ \vdots & \vdots & \ddots &
\end{matrix}\right)_{i, j} \ ,
\label{eq:unitarytransform}
\end{equation}
and to define new operators $A_{i,\omega}(t)$ given by
\begin{equation}
L_{\alpha, \omega}(t) = \sum_{i}u_{i,\alpha}(\omega) A_{i,\omega}(t) \ .
\label{eq:L}
\end{equation}
In this basis, we can write the dissipative part in diagonal form as:
\begin{eqnarray}
\label{eqt:dissdiag}
{\mathcal{L}}_{\textrm{WCL}}  [\rho(t)]  & = &\sum_{i}\sum_{\omega}\gamma'_i(\omega)\left(
A_{i,\omega}(t) \rho(t) A^\dagger_{i,\omega}(t)\phantom{\frac{1}{2}} \right. \notag \\
&& \hspace{-0.5cm} \left. \qquad\qquad - \frac{1}{2} \left\{ A_{i,\omega}^{\dagger}(t) A_{i,\omega}(t) , \rho(t) \right\} \right) \ .
\end{eqnarray}

\section{Stochastic Schr\"odinger Equation}
\label{sec:III}
With Eq.~\eqref{eqt:dissdiag}, the master equation Eq.~\eqref{eqt:ME2-H} is in diagonal form and can be unravelled into quantum trajectories. The trajectory is described by a stochastic differential equation (SDE) in the form of jumps or diffusion. Let us consider the case where the coefficients $\gamma'_i(\omega)$ in Eq.~\eqref{eqt:dissdiag} also depend on time. If all $\gamma'_i(\omega, t) \geq 0$, then the dynamics is completely positive (CP)-divisible~\cite{laine2010measure}, and the master equation can be unravelled by using the known unravelling of the the time-independent SDE case~\cite{Breuer:2002,gardiner2004quantum,brun:719}, simply by replacing the time-independent operators and coefficients by the time-dependent ones.

Such an unravelling is also possible, but with modifications, when the dynamics is positive (P)-divisible, i.e., where $\gamma'_i(\omega, t)$ need not be all positive.\footnote{The condition on $\gamma'_i(\omega, t)$ such that the map is P-divisible can be found in the proof given in~\cite{breuer2009stochastic} or Eq.~(25) in~\cite{Caiaffa:2017aa}.} This can be in the form of:
\begin{itemize}
\item {Jump} trajectories: the master equation is unravelled via the non-Markovian quantum jump method (NMQJ)~\cite{piilo2008non, breuer2009stochastic,piilo2009open,harkonen2010jump}, where terms with negative coefficients $\gamma'_i(\omega, t)$ describe the negative channel.
\item {Diffusive} trajectories: recent work on diffusive trajectories~\cite{Caiaffa:2017aa} replaces $\gamma'_i(\omega, t)$ and the operators by the eigenvalues and eigenvectors of a positive transition rate operator $W$ (Eq.~(11) in~\cite{Caiaffa:2017aa}). P-divisible dynamics can be unravelled into a SDE in terms of such eigenvalues and eigenvectors.
\end{itemize}

In the following, we focus on the case of CP maps [with all $\gamma'_i(\omega)\geq 0$] and unravel the master equation in the quantum jumps picture.

\subsection{Unravelling the master equation}
First we absorb the $\gamma'$ coefficients into the definition of $A_i$:
\beq 
\sqrt{\gamma'_i(\omega)}A_{i,\omega}(t)\rightarrow A_i(t) \ .
\label{eqt:Aired}
\eeq
In this redefinition, the index $i$ now includes the Bohr frequencies.
We write Eq.~\eqref{eqt:ME2-H} in terms of an effective non-Hermitian Hamiltonian $H_{\eff}$:
\begin{align}
\label{eqt:effdiagonalform}
\frac{d}{dt}{\rho}(t) &= - i\left(H_{\text{eff}}(t)\rho_S(t) - \rho_S(t)H^{\dagger}_{\text{eff}}(t)\right) \nonumber\\
&\phantom{{}=}+ \sum_{i}A_{i}(t) \rho_S(t) A^\dagger_{i}(t)   \,, 
\end{align}
where 
\begin{equation}
\label{eq:Heff}
H_{\text{eff}}(t) = H_{\textrm{S}}(t) + H_{\text{\textrm{LS}}}(t) - \frac{i}{2} \sum\limits_{i}{A}_{i}^{\dagger}(t) {A}_{i}(t)\,.
\end{equation}%
Equation~\eqref{eqt:effdiagonalform} can be unravelled into quantum trajectories in the quantum jumps picture, where each trajectory describes the stochastic evolution of a pure state (if the initial state $\rho$ is mixed, $\rho = \sum_i p_i \ketbra{\psi_i}$, then the evolution can be performed on each initial pure state).  The stochastic evolution of the pure state can be written in terms of a stochastic Schr\"{o}dinger equation (in It$\hat{\text{o}}$ form), the ensemble average of which is equivalent to the master equation:
\begin{align}
\label{eqt:sse}
d\ket{\psi(t)} &= \left(- i{H}_{\text{eff}}(t) + \frac{1}{2}\sum_{i}\braket{{A}_{i}^{\dagger}(t){A}_{i}(t)}\right) dt \ket{\psi(t)} \nonumber \\
&\phantom{{}=}+ \sum_{i} dN_{i}(t)\left(  \frac{{A}_{i}(t)}{\sqrt{\braket{{A}^{\dagger}_{i}(t){A}_{i}(t)}}} - \mathds{1}        \right)\ket{\psi(t)}
\end{align}
where 
\beq
\braket{{A}_{i}^{\dagger}(t){A}_{i}(t)} \equiv \braket{\psi(t)|{A}_{i}^{\dagger}(t){A}_{i}(t)|\psi(t)} = \lVert{A}_{i}(t)\ket{\psi(t)}\rVert^2\ .
\label{eq:12}
\eeq 
We give a derivation of Eq.~\eqref{eqt:sse} below.
The first term on the r.h.s. of Eq.~\eqref{eqt:sse} gives a deterministic evolution composed of a Hermitian component [$-i(H_{\text{S}}(t) + H_{\text{\textrm{LS}}}(t))$] and a ``drift'' component 
\beq
D(t) \equiv  \frac{1}{2}\sum_{i} A_{i}^{\dagger}(t)A_{i}(t) - \braket{A_{i}^{\dagger}(t)A_{i}(t)} \ ,
\eeq
and the second term describes the stochastic jump process.  The stochastic variable $dN_{i}(t)\equiv N_i(t+dt) - N_i(t)$ is the number of jumps of type $i$ in the interval $dt$, where we have denoted by $N_i(t)$ the number of jumps of type $i$ up to time $t$.  The expectation value of the stochastic variable is given by  \cite{Breuer:2002}:
\beq
E[dN_{i}(t)] = \braket{{A}_{i}^{\dagger}(t){A}_{i}(t)}dt \,.
\eeq
Since the probability of a jump occurring scales linearly with $dt$, the probability of having more than one jump vanishes faster than $dt$, so as $dt\to 0$ only one jump out of all possible types during $dt$ is permitted. Therefore we can write~\cite{gardiner2004quantum}:
\begin{equation}
\label{eqt:stochastic1}
dN_{i}(t) = \left\{
                \begin{array}{ll}
                  1 &\text{with prob.                } \braket{{A}_{i}^{\dagger}(t){A}_{i}(t)}dt\\
                  0 &\text{with prob.                }  1 - \braket{{A}_{i}^{\dagger}(t){A}_{i}(t)}dt
                \end{array}
\right.
\end{equation}
with the It$\hat{\text{o}}$ table:
\bes
\label{eqt:stochastic2}
\begin{align}
dN_{j}(t)dN_{k}(t) &= \delta_{jk}dN_{j}(t)  \\
dN_{j}(t)dt &= 0 \,.
\end{align}
\ees
From Eq.~\eqref{eqt:stochastic1}, the probability of any jump occurring, $\sum_{i}\braket{{A}_{i}^{\dagger}(t){A}_{i}(t)}dt$, is small compared to the probability of no jump occurring, so $\sum_{i}dN_{i}(t) = 0$ most of the time. During the infinitesimal time-step $dt$, if $\sum_{i}dN_{i}(t) = 0$, then only the deterministic evolution takes places; if however $\sum_{i}dN_{i}(t) = 1$, then a jump occurs. When a jump occurs, it dominates over the deterministic evolution, which is proportional to $d t$, and the deterministic part can be ignored.

\subsection{Deterministic evolution and jump process}
\label{ssec:A}
\begin{figure}[t!]
\centering
\includegraphics[width=3in]{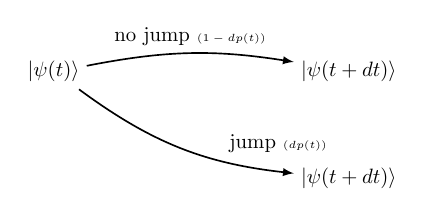}  
\caption{A depiction of the stochastic evolution of the state $\ket{\psi}$ by an infinitesimal time state at time $t$.}  
\label{fig:branch}
\end{figure}

We now derive Eq.~\eqref{eqt:sse} by explaining how each probability element appears.  Let us denote by $\ket{\psi(t)}$ and $\ket{\tilde{\psi}(t)}$ the normalized and unnormalized state vectors respectively, and assume they are equal at time $t$, i.e., $\ket{\tilde{\psi}(t)} = \ket{\psi(t)}$.  
For the infinitesimal time-step from $t$ to $t+dt$, the state vector evolution from $\ket{\psi(t)}$ to $\ket{\psi(t+dt)}$ involves two possibilities: either no jump occurring (with probability $1- dp$) or a jump occurring (with probability $dp$).  This is depicted in Fig.~\ref{fig:branch}.

When no jump occurs, the evolution is described by the Schr\"{o}dinger equation associated with $H_\eff$, and since the effective Hamiltonian is non-Hermitian the norm of the state vector is not preserved during the evolution:
\begin{equation}
\label{eqt:Schr}
\frac{d\ket{\tilde{\psi}(t)}}{dt} = -iH_{\text{eff}}(t) \ket{\tilde{\psi}(t)} \,.
\end{equation}
The resulting state after one infinitesimal time-step $dt$ is:
\bes
\label{eqt:detun}
\begin{align}
\ket{\tilde{\psi}(t+dt)} &= \exp\left[-iH_{\text{eff}}(t)dt\right]\ket{\psi(t)} \\   
&= \left[\mathds{I} - i dt H_{\text{eff}}(t)   + O(dt^2)\right] \ket{\psi(t)} \ .
\end{align}
\ees
The norm squared $\|\tilde{\psi}(t+dt)\|^2$ is the probability of the conditional evolution under $H_\eff$, 
so that (as we show explicitly in Appendix~\ref{app:A}) the jump probability is given by
\beq
1-\|\tilde{\psi}(t+dt)\|^2 = dt  \sum_{i}\braket{{A}_i^{\dagger}(t){A}_i(t)} + O(dt^2) 
\label{eqt:squareofnorm}
\eeq
[recall Eq.~\eqref{eq:12} and note that $H_{\text{\textrm{S}}}(t)+H_{\text{\textrm{LS}}}(t)$ cancels out to first order].
Therefore we can identify the infinitesimal jump probability $dp(t)$ with the r.h.s. of Eq.~\eqref{eqt:squareofnorm}, i.e., to first order in $dt$:
\bes
\label{eq:jumprate}
\begin{align}
\label{eqt:jumpprob}
dp(t) &= \sum_{i} dp_i (t) = dt \lambda(t) \\
\label{eqt:typeprob}
dp_i(t) &= dt \braket{{A}_i^{\dagger}(t){A}_i(t)} \ ,
\end{align}
\ees
where $\lambda(t) = \dot{p}(t)$ is the jump rate, and $dp_i(t)$ is the probability of the jump of type $i$.  Note that since our definition of the $A_i$ operators includes the rates $\gamma'$ [recall Eq.~\eqref{eqt:Aired}], the jump rate depends on the instantaneous Bohr frequencies and the KMS condition.

When the jump of type $i$ occurs the state is updated as:
\begin{equation}
\label{eqt:jumpun}
\ket{\tilde{\psi}(t+dt)} ={A}_i(t)\ket{\psi(t)} \,.
\end{equation}

We can unify the two possibilities in Eq.~\eqref{eqt:detun} and Eq.~\eqref{eqt:jumpun} as a stochastic Schr\"{o}dinger equation for the unnormalized state vector where only terms of order $dt$ are kept:
\bes
\label{eqt:sse2}
\begin{align}
d\ket{\tilde{\psi}(t)} &= \ket{\tilde{\psi}(t + dt)} - \ket{\tilde{\psi}(t)} \\
&= -i dt H_{\text{eff}}(t)\ket{\tilde{\psi}(t)}  + \nonumber\\
&\phantom{=}\sum_{i}dN_{i}(t)\left({A}_i(t)-\mathds{1}\right)\ket{\tilde{\psi}(t)} \,,
\end{align}
\ees
where we used $\ket{\tilde{\psi}(t)} = \ket{\psi(t)}$. The stochastic element $dN_{i}(t)$ has the properties given in Eqs.~\eqref{eqt:stochastic1} and~\eqref{eqt:stochastic2}; $\mathds{1}$ is subtracted since when the jump  occurs $\sum_{i}dN_{i}(t)\ket{\tilde{\psi}(t)} = \ket{\tilde{\psi}(t)}$ and the term involving $H_{\text{eff}}(t)$ is absent, so in this manner we ensure that $\ket{\tilde{\psi}(t)}$ is appropriately subtracted from the r.h.s.

We can write a similar expression for the normalized state vector $\ket{\psi(t)}$ by normalizing Eqs.~\eqref{eqt:detun} and~\eqref{eqt:jumpun}.  If a deterministic evolution occurs, we have
\bes
\label{eqt:deterministicequivalence}
\begin{align}
\label{eqt:deterministicequivalence-1}
\ket{\psi(t + dt)} &= \frac{\exp\left[-iH_{\text{eff}}(t)dt\right]\ket{\psi(t)}}{\norm{\exp\left[-iH_{\text{eff}}(t)dt\right]\ket{\psi(t)}}} \\
&\hspace{-1cm}= \frac{\left(\mathds{1} - i dt H_{\text{eff}}(t)+O(dt^2)\right)\ket{\psi(t)}}{\sqrt{1 - dt \sum_{i}\braket{{A}_i^{\dagger}(t){A}_i(t) } + O(dt^2)}}\\
&\hspace{-1cm} = \left(\mathds{1} - i dt H_{\text{eff}}(t) + \frac{1}{2}\sum_{i} \braket{{A}_i^{\dagger}(t){A}_i(t)}dt \right)\ket{\psi(t)}  \notag \\
& + {O}(dt^2)\ .
\end{align}
\ees
If a jump of type $i$ occurs, we have
\begin{equation}
\ket{\psi(t + dt)} = \frac{{A}_i(t)\ket{\psi(t)}}{\norm{{A}_i(t)\ket{\psi(t)}}} = \frac{{A}_i(t)\ket{\psi(t)}}{\sqrt{\braket{{A}^{\dagger}_{i}(t){A}_{i}(t)}}}\,.
\label{eqt:jump}
\end{equation}
Therefore, in analogy to Eq.~\eqref{eqt:sse2} we can write the stochastic Schr\"odinger equation for the normalized state as in Eq.~\eqref{eqt:sse}.

\section{Simulation procedure for adiabatic quantum trajectories}
\label{sec:IV}
In this section we formulate an algorithm for implementing adiabatic quantum trajectories. We start by noticing that the update in Eq.~\eqref{eqt:deterministicequivalence-1} corresponds to the evolution by the first part of the stochastic Schr\"{o}dinger equation. Therefore, the deterministic evolution in the first term of Eq.~\eqref{eqt:sse} is equivalent to propagating the state vector via the Schr\"{o}dinger equation with $H_\eff(t)$ and then renormalizing it. 

When a jump occurs, one of the operators $A_i(t)$ is applied.  The relative weight of each $A_i(t)$ is $dp_i(t)$, given in Eq.~\eqref{eqt:typeprob}. In this case, the state is evolved as in Eq.~\eqref{eqt:jumpun}, and the normalized state is given in Eq.~\eqref{eqt:jump}. 
The update in Eq.~\eqref{eqt:jump} corresponds to the evolution by the second term of Eq.~\eqref{eqt:sse}. 

This provides a direct way to algorithmically implement the quantum trajectories method.   Starting from a known normalized initial state, the state is evolved via a sequence of deterministic evolutions and jumps, as in Eqs.~\eqref{eqt:detun} and~\eqref{eqt:jump}, by drawing a random number at each finite but small time-step $\Delta t$ and determining which of the two choices to take.  Compared to the standard time-independent case, the size of the time-step must satisfy additional conditions in order for the approximations to hold:
\beq
\Delta t \ll \min_t \left\{ \frac{2\vertiii{H_{\text{eff}}(t)}}{\| \dot{H}_{\text{eff}}(t)\|},
\frac{1}{\vertiii{H_{\text{eff}}(t)}},
\left|\frac{ \lambda(t) }{ \lambda^2(t)-\dot{\lambda}(t)}\right| \right\} \ ,
\label{eq:newcond}
\eeq
where $\|\cdot\|$ is the operator norm (largest singular value).  We give a proof of this new bound in Appendix~\ref{app:B}.  While the second and third terms reduce to the known conditions for the time-independent case, the first term in Eq.~\eqref{eq:newcond} is unique to the time-dependent case and reflects the error associated with time evolution under a time-dependent effective Hamiltonian.  This term highlights the fact that the faster the effective Hamiltonian and its eigenstates vary in time, the smaller is the time-step required to properly follow the trajectory. Eq.~\eqref{eq:newcond} can also be viewed as the physical timestep upper bound in weak measurement.

However, drawing a random number at each time-step is computationally expensive, 
so it is more efficient to use the waiting time distribution~\cite{Breuer:2002}
to determine the first jump event.  As we mentioned before [Eq.~\eqref{eqt:squareofnorm}] the square norm of the unnormalized wavefunction at $t + dt$ gives the probability of no jump during the infinitesimal interval $[t,t+dt]$. We show in Appendix~\ref{sec:C} that starting from the normalized state $\ket{\psi(t)}$, the probability of no jump occurring in the finite (not necessarily small) time interval $[t,t+\tau]$ is given by
\begin{equation}
\|\tilde{\psi}(t+\tau)\|^2 =  \exp\left(-\int_t^{t+\tau}\lambda(s)ds \right) \,,
\end{equation}
where the jump rate $\lambda(t)$ is given in Eq.~\eqref{eqt:jumpprob}. 
With this, the simulation procedure for one single trajectory is as follows, starting from $t$:
\begin{itemize}
\label{item:algo}
\item Draw a random number $r$.
\item Propagate the unnormalized wavefunction by solving the Schr\"{o}dinger equation with $H_\eff$ [Eq.~\eqref{eqt:Schr}] until the jump condition is reached at $t+\tau$, i.e., for $\tau$ such that $\braket{\tilde{\psi}(t+\tau)|\tilde{\psi}(t+\tau)} \leq  r$. (Recall that the norm of the unnormalized wavefunction will keep decreasing in this process.)
\item Determine which jump occurs by drawing another random number and update the wavefunction by applying jump operators, and renormalize.
\item Repeat the above steps with the new normalized state.  
\item Repeat until the final simulation time is reached.
\end{itemize}

We prove that averaging over quantum trajectories recovers the master equation in Appendix~\ref{app:B}. Specifically, we show there that if we denote the state of the $k$-th trajectory at time $t$ by $\ket{\psi_k(t)}$, then we can approximate the master equation solution for the density matrix $\rho(t)$ as $\frac{1}{n}\sum_{k=1}^{n}\ketbra{\psi_k(t)}$ for large $n$. Choosing a basis $\{\ket{z_i}\}$ for the system Hilbert space, we can thus approximate the density matrix element $\bra{z_i} \rho(t) \ket{z_j}$ as $\frac{1}{n}\sum_{k=1}^{n}\langle{z_i} \ketbra{\psi_k(t)} {z_j}\rangle$ for large $n$.
%

\section{Case studies}
\label{sec:V}
We consider a system of $N$ qubits with a transverse-field Ising Hamiltonian given by
\begin{subequations}
\label{eqt:H_S}
\begin{eqnarray}
\label{eqt:H_Sa}
{H_{\textrm{S}}(t)} &=& A(t) H_{\textrm{S}}^X + B(t) H_{\textrm{S}}^Z {\ , } \\
H_{\textrm{S}}^X &\equiv& -\sum_{i=1}^N \sigma_i^x { \ ,}\\
H_{\textrm{S}}^Z &\equiv&  - \sum_{i=1}^N h_i
\sigma_i^z + \sum_{i>j=1}^N  J_{i j} \sigma_i^z \sigma_j^z \ .
\end{eqnarray}
\end{subequations}

\begin{figure}[t]
\centering
\subfigure[\ ]{\includegraphics[width=0.49\textwidth]{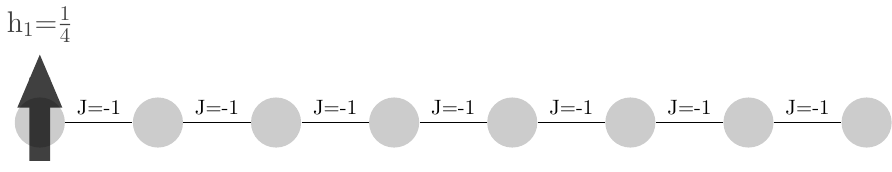}\label{fig:4a}} 
\subfigure[\ ]{\includegraphics[width=0.15\textwidth]{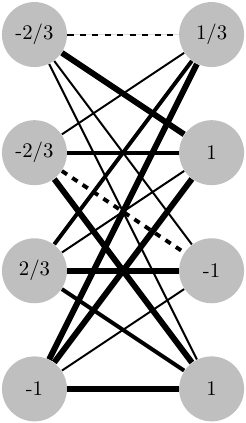}\label{fig:4b}} 
\subfigure[\ ]{\includegraphics[width=0.49\textwidth]{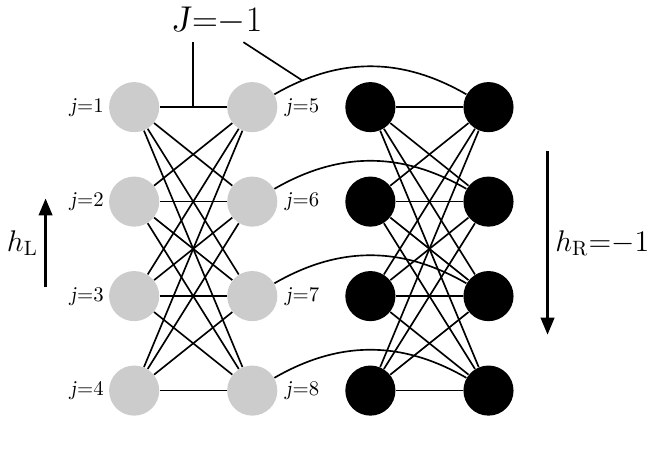}\label{fig:4c}} 
\caption{Graphs of (a) the $8$-qubit chain, (b) the $8$-qubit Hamiltonian exhibiting a small gap, and (c) the $16$-qubit ``tunneling-probe" Hamiltonian of Ref.~\cite{Boixo:2014yu}. (a) Only the first qubit is subjected to an applied field and each qubit is ferromagnetically coupled with $J = -1$. (b)  Solid lines corresponds to ferromagnetic coupling and dashed lines corresponds to  antiferromagnetic coupling.  The thickness denotes the strength of the coupling.  Local fields are shown inside the circles.  Full parameters are given in Eq.~\eqref{eqt:Case0989}.  (c) The left $8$-qubit cell and right $8$-qubit cell are each subjected to applied fields with opposite direction. Each qubit is ferromagnetically coupled to others as shown by the lines, with $J = -1$.}
\end{figure}

\begin{figure}[t]
\begin{center}
{\includegraphics[width=0.49\textwidth]{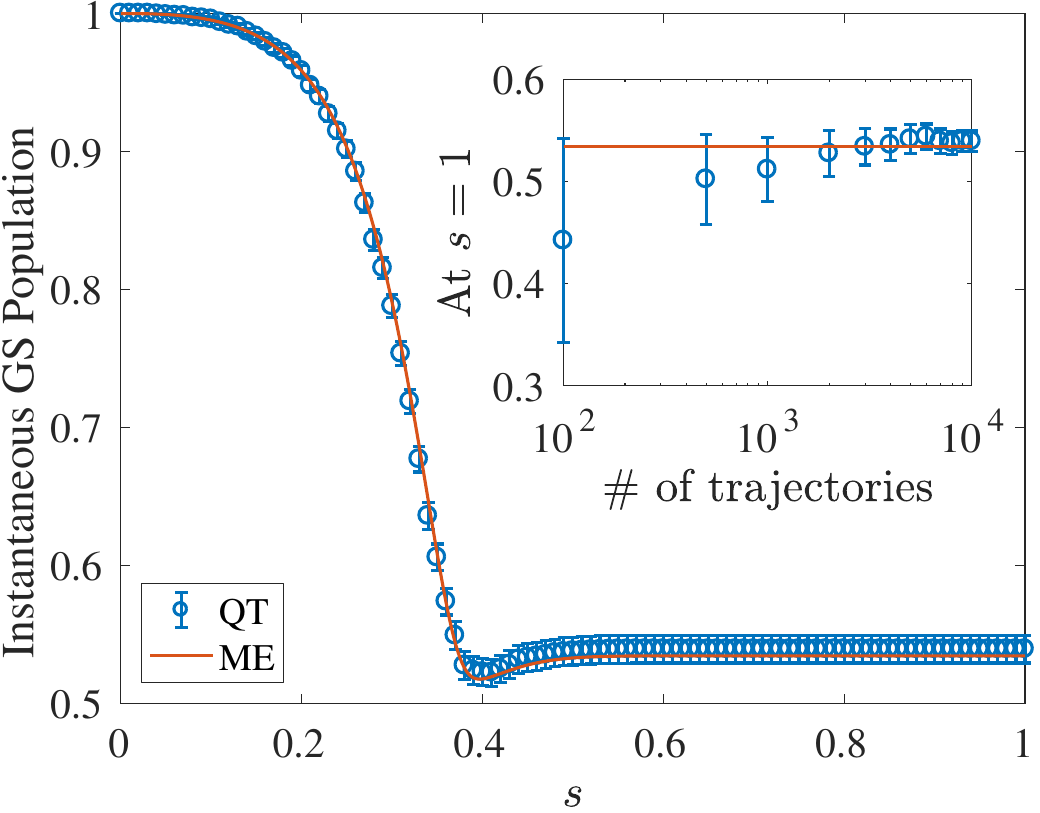}} 
\end{center}
\caption{The evolution of the population in the instantaneous ground state for the $8$-qubit problem in Eqs.~\eqref{eqt:H_S} and \eqref{eqt:Ising} for a total time of $t_f = 10 \mu s$ and temperature $2.6$GHz (in $\hbar\equiv 1$ units), as a function of the normalized time $s=t/t_f$.  The quantum trajectories results with $10^4$ trajectories (QT, blue circles) are in excellent agreement with the adiabatic master equation (ME, red solid line).   Inset: the convergence of the ground state population (averaged over quantum trajectories) towards the master equation result as a function of the number of trajectories,  at $s = 1$.  The error bars represent $2 \sigma$ confidence intervals, where $\sigma$ is the standard deviation of the mean generated by taking $10^3$ bootstraps over the number of trajectories. }
   \label{fig:3qubits8qubits}
\end{figure}

We assume that the qubit-system is coupled to independent, identical bosonic baths, with the bath and interaction Hamiltonian being 
\bes
\begin{align}
\label{eq:SBm}
H_B &= \sum_{i=1}^N \sum_{k=1}^\infty \omega_k b_{k,i}^\dagger b_{k,i} \ ,  \\
 H_{I} &= g \sum_{i=1}^N
\sigma_i^z \otimes \sum_k \left(b_{k,i}^\dagger + b_{k,i} \right) \ ,
\end{align}
\ees
where $b_{k,i}^\dagger$ and $b_{k,i}$ are, respectively, raising and lowering operators for the $k$-th
oscillator mode with natural frequency $\omega_k$. 
The bath correlation functions appearing in Eq.~\eqref{eqt:ME2-L} are given by
\begin{equation}
\gamma_{ij}(\omega) = 2\pi g^2\frac{\omega e^{-|\omega|/\omega_c}}{1-e^{-\b\omega}} \delta_{ij} \ ,
\end{equation}
arising from an Ohmic spectral density~\cite{ABLZ:12-SI}, and satisfy the KMS condition~\eqref{eq:KMS}.

\subsection{$8$-qubit chain} \label{sec:8qubitchain}
As a first illustrative example and as a consistency check, we reproduce the master equation evolution of the $8$-qubit ferromagnetic Ising spin chain in a transverse field studied in Ref.~\cite{ABLZ:12-SI}. For this problem, the Ising parameters are given by [also shown in Fig.~\ref{fig:4a}]
\begin{equation} \label{eqt:Ising}
 h_1 = \frac{1}{4} \ , \quad h_{i>1}=0  \ , \quad  J_{i, i+1} = -1 \ , \quad i = 1, \dots, 8 \ .
\end{equation}
The functional form of the functions $A(t)$ and $B(t)$ is given in Appendix \ref{app:Chain} (they are the annealing schedule of the D-Wave One ``Rainier" processor, described in detail, e.g., in Ref.~\cite{q108}, and were also used in the original AME study of the $8$-qubit chain~\cite{ABLZ:12-SI}). 
The effect of changing the schedule is small; for comparison we provide in Appendix~\ref{app:Chain} results for a linear annealing schedule with the same bath parameters. As shown in Fig.~\ref{fig:3qubits8qubits}, we recover the master equation solution within the error bars.  The initial state is the ground state of $H_{S}(0)$, which is the uniform superposition state.  

\begin{figure}[t] 
\centering
\includegraphics[width=0.49\textwidth]{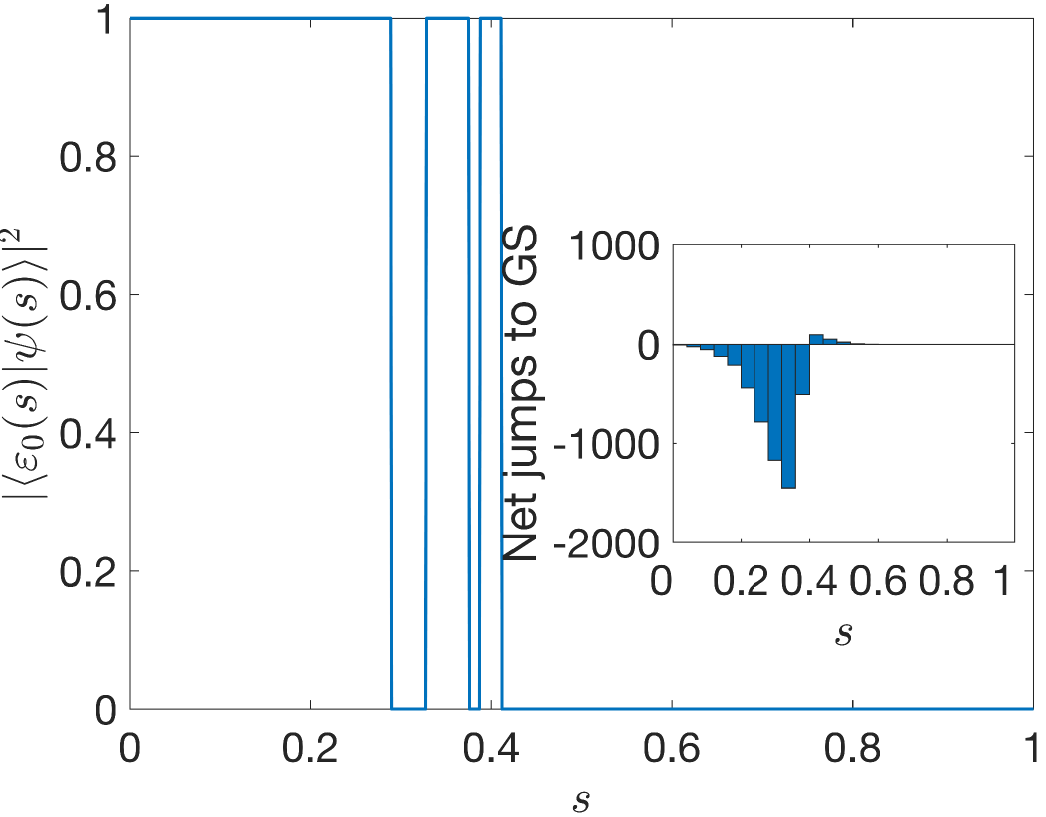}
\caption{The overlap squared of the (normalized) state with the instantaneous ground state of $H_{S}(t)$ for a typical single trajectory of the 8 qubit chain in Sec.~\ref{sec:8qubitchain}, with $t_f  = 10\mu s$ and temperature $2.62$GHz, as a function of the normalized time $s=t/t_f$. The sudden changes in the overlap are due to the action of the jump operators $\left\{A_i(t) \right\}$, taking the state from one eigenstate to another. This is to be contrasted with the smooth behavior of Fig.~\ref{fig:3qubits8qubits} when we average over different trajectories.  Inset: a histogram of the net number of jumps to the instantaneous ground state (GS).  A negative number indicates a jump out of the ground state, and a positive number indicates a jump towards the ground state. The change from negative to positive net jumps occurs at the minimum gap point.}
\label{fig:8qubitssingle}
\end{figure}

It is illustrative to see how a single trajectory differs from the averaged case, and we show this in Fig.~\ref{fig:8qubitssingle}.  Instead of the smooth change in the population as observed in the averaged case, the single trajectory behaves like a step-function. This is explained by the fact that the drift term vanishes if $\ket{\psi(t)}$ is a nondegenerate eigenstate, as shown in Appendix~\ref{app:eigenstate-proof}. Therefore changes in the state's overlap with the instantaneous ground state occur only due to the jump operators.  In this picture, the ground state population revival observed after the minimum gap is crossed is associated with jumps from the first excited state (or higher states for large $T$) to the ground state. After the minimum gap, there are more transitions back to the ground state than out of the ground state (see the inset of Fig.~\ref{fig:8qubitssingle}). Such a difference (divided by the number of trajectories) leads to the rise of the ground state population. 

Using Eq.~\eqref{eq:jumprate}, we can give an explicit expression for the jump rate from the first excited state back to ground state. As shown in  Appendix~\ref{app:lindbladfermi}, this is given by:
\begin{equation}
\label{eq:inctunnrate}
\lambda_{1\rightarrow0}(t) = 
\sum_{\alpha = 1}^{8} \gamma_\alpha(\omega_{10})|\langle\psi_0(t)|\sigma^{z}_{\alpha} |\psi_1(t)\rangle|^2\ .
\end{equation}

\begin{figure}[t]
\begin{center}
{\includegraphics[width=0.49\textwidth]{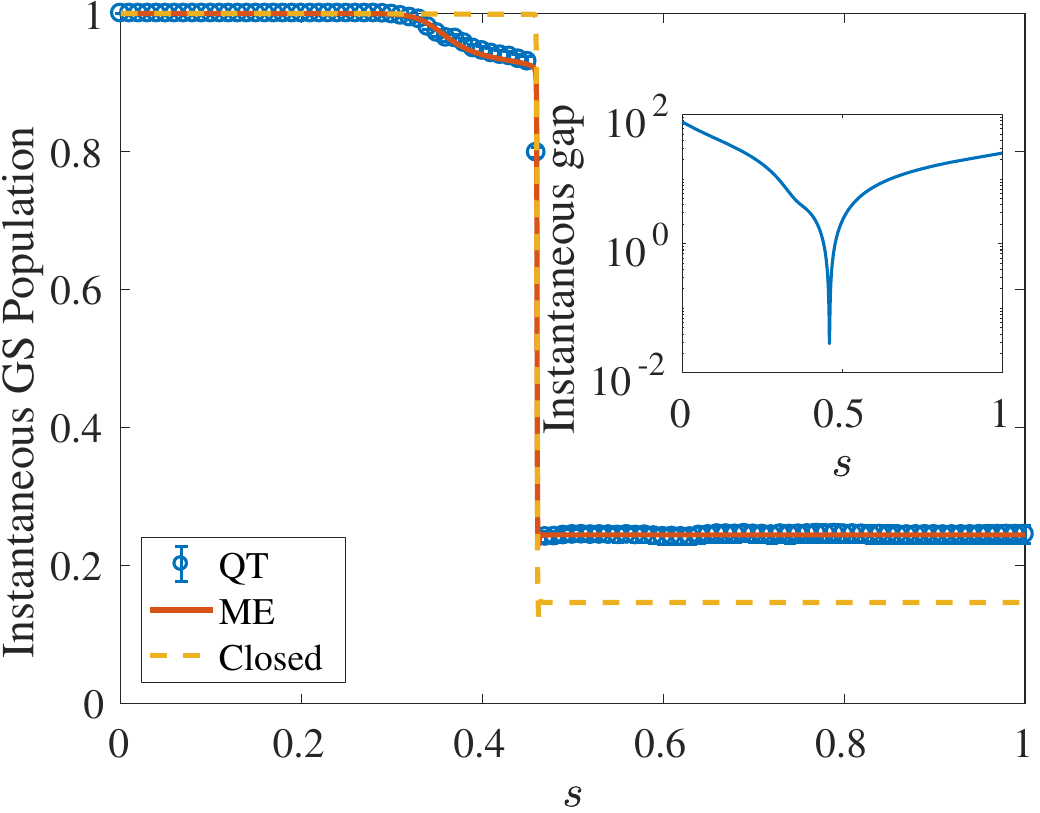}} 
\end{center}
\caption{The evolution of the population in the instantaneous ground state for the $8$-qubit problem in Eqs.~\eqref{eqt:H_S} and \eqref{eqt:Case0989} for a total time of $t_f = 10 \mu s$ and temperature $1.57$GHz, as a function of the normalized time $s=t/t_f$.  The quantum trajectories results with $5\times 10^3$ trajectories (blue circles) are in excellent agreement with the master equation (red solid line). We also show the closed-system evolution (yellow dashed line) to highlight that the evolution is not adiabatic.   Inset: the instantaneous energy gap between the first excited state and the ground state during the anneal. The minimum occurs at $s^*=0.46$, coinciding with the sharp discontinuity observed in the instantaneous ground state population.}
   \label{fig:case0989}
\end{figure}

\subsection{$8$-qubit non-adiabatic example} 
\label{sec:case0989}

We now consider an $8$-qubit problem with a sufficiently small minimum gap such that the closed-system evolution is not adiabatic even with $t_f = 10 \mu s$ and using the DW2X annealing schedule (described in detail, e.g., in Ref.~\cite{Albash:2017aa}).  While this strictly violates the assumptions under which the AME is derived,%
\footnote{Equation~(27) in Ref.~\cite{ABLZ:12-SI} is a necessary condition for the validity of the AME. It states that $\frac{h}{\Delta^2 t_f}\ll 1$, where $\Delta$ is the ground state gap and $h = \max_{s\in [0,1];a,b} |\bra{\epsilon_a(s)}|\partial_sH_S(s)\ket{\epsilon_b(s)}|$, with $s=t/t_f$ and $\ket{\epsilon_a(s)}$ being the instantaneous $a$-th eigenstate of the system Hamiltonian $H_S(s)$. We find that for $t_f = 10 \mu s$, the l.h.s.$\approx 5$.}
we can ask about the dynamics associated with the master equation irrespective of its origins. We are interested in this example since it illustrates some aspects of the quantum trajectories picture which are not visible in the adiabatic limit, as explained below.

The Ising Hamiltonian $H_S^Z$ is defined with parameters:
\bes \label{eqt:Case0989}
\begin{align}
3 \vec{h} &= (-2, -2, 2, -3, 1, 3, -3, 3) \ , \\
 3J_{0,4} &= \phantom{-}1, 3J_{0,5} = -3,  3J_{0,6} = -1,  3J_{0,7} = -1,  \\
 3J_{1,4} &= -1, 3J_{1,5} = -2,  3J_{1,6} = \phantom{-}2,  3J_{1,7} = -3,  \\
 3J_{2,4} &= -2, 3J_{2,5} = -1,  3J_{2,6} = -3,  3J_{2,7} = -2, \\
 3J_{3,4} &= -3, 3J_{3,5} = -3,  3J_{3,6} = -1,  3J_{3,7} = -3, 
\end{align}
\ees
Figure~\ref{fig:case0989} shows our simulation results, obtained by solving the AME directly and by using the trajectories approach. Reassuringly, the agreement between the two is excellent. Also plotted are the closed system results for this problem, which exhibit a sharp diabatic transition out of the ground state at the minimum gap point (the small gap is shown in the inset). The AME and trajectories results show that the ground state population loss starts before the diabatic transition, due to thermal excitations, but that the ground state population loss is partially mitigated by the presence of the thermal bath, with the open system ending up with a higher ground state population than the closed system. 

\begin{figure}[t] 
\centering
\includegraphics[width=0.49\textwidth]{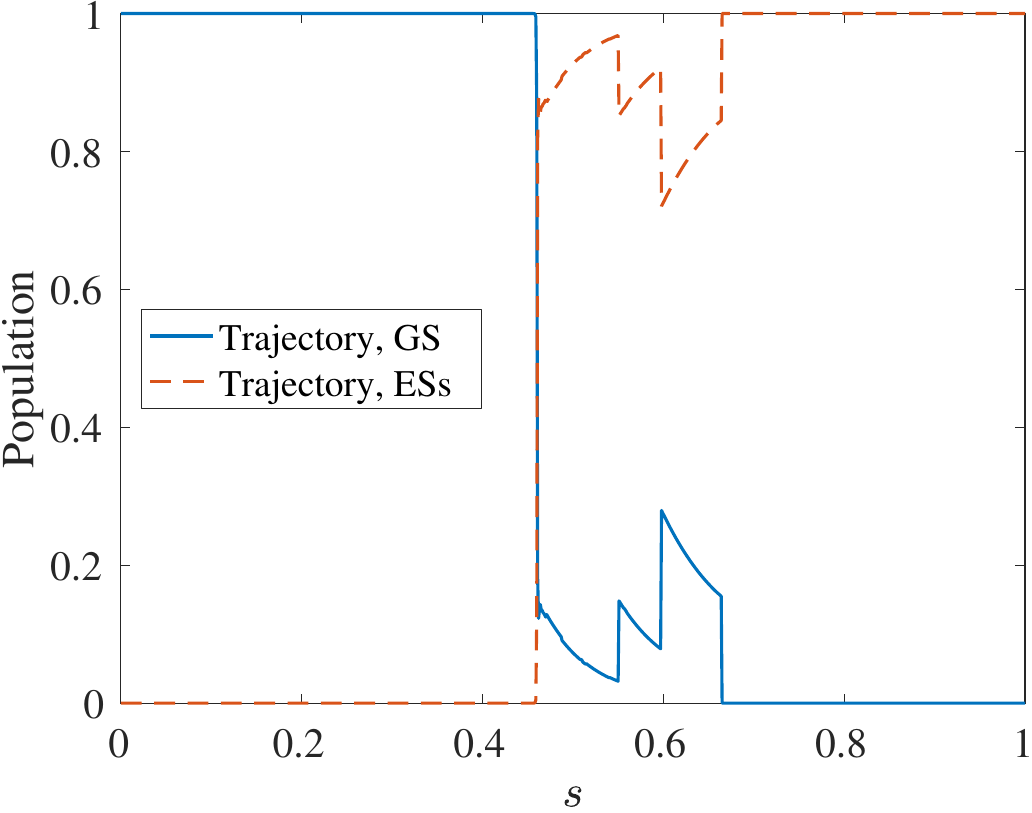}
\caption{The overlap squared of the (normalized) state with the instantaneous ground state of $H_{S}(t)$ (blue solid curve) and the sum of the first three instantaneous excited states (red dashed curve) for a typical single trajectory of the $8$-qubit problem in Sec.~\ref{sec:case0989}, with $t_f  = 10\mu s$ and temperature $1.57$GHz, as a function of the normalized time $s=t/t_f$. A diabatic transition occurs at $s^* = 0.46$. The continuous decay immediately afterward and between the `incomplete' jumps is to be contrasted with the step-function like single trajectories of the adiabatic case seen in Fig.~\ref{fig:8qubitssingle}. 
}
\label{fig:Case0989single}
\end{figure}

The diabatic transition results in different trajectories than those observed for the adiabatic case in Sec.~\ref{sec:8qubitchain}.  We show such a case in Fig.~\ref{fig:Case0989single}.  Instead of the pulse-like structure seen in Fig.~\ref{fig:8qubitssingle}, we observe a combination of both drifts and jumps.  Because the diabatic transition generates a non-eigenstate that is a coherent superposition of the ground state and first excited state, drifts caused by the environment show up in the subsequent evolution.  Furthermore, this superposition also means that the Lindblad operator associated with $\omega = 0$, if having different component weights in Eq.~\eqref{eq:Lindblad2}, can also induce jumps (e.g., the jumps around $s = 0.6$ in Fig.~\ref{fig:Case0989single}), an effect that is completely absent in the adiabatic case.  These jumps need not project the state completely onto an instantaneous energy eigenbasis state, but they can change the relative weights on the different occupied eigenstates, which manifest themselves as `incomplete' jumps in the trajectories.

We reemphasize that due to the violation of the conditions under which the AME is derived, the observations we have reported for this example are strictly valid only when the AME is taken at face value, and do not necessarily reflect actual physical dynamics.

\begin{figure}[t] 
   \centering
   \includegraphics[width=0.49\textwidth]{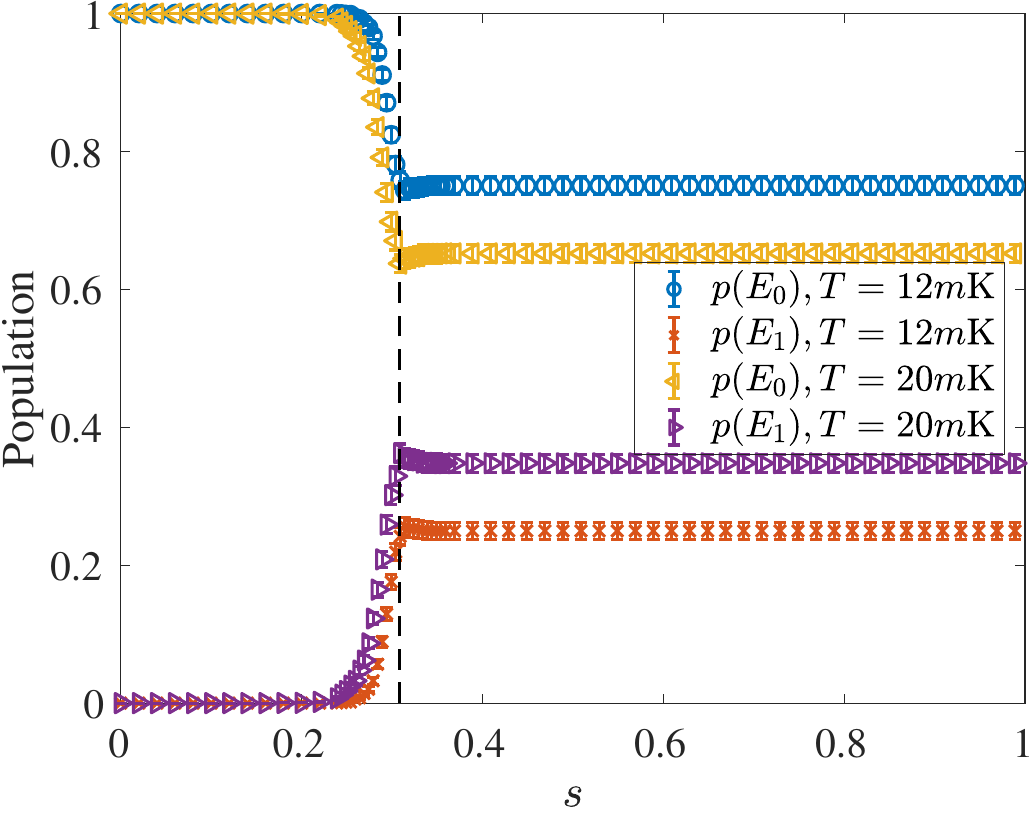} 
   \caption{Quantum trajectory results for a temperature of $12m$K and $20m$K using the DW2X annealing schedule. Results are averaged over $5000$ trajectories. A revival of instantaneous ground state population occurs after the minimum gap (shown by the dashed vertical line).}
   \label{fig:GoogleGadget}
\end{figure}
%

\subsection{$16$-qubit ``tunneling-probe" Hamiltonian}
\label{sec:16qubit}

In order to demonstrate the computational utility of the trajectories approach over the master equation approach, we now give results for a $16$-qubit system first studied in Ref.~\cite{Boixo:2014yu} for the purpose of probing tunneling in quantum annealing. 

For this problem, the parameters of Eq.~\eqref{eqt:H_S} are [also shown in Fig.~\ref{fig:4b}]:
\begin{equation} 
\label{eqt:H_S16}
 h_{\text{L}} = 0.44\ , \quad h_{\text{R}} = -1  \ , \quad  J_{i, i+1} = -1 \ , \quad i = 1, \dots, 16 \ .
\end{equation}
where the sets $L$ and $R$ range over $i = 1,\dots, 8$ and $i = 9,\dots, 16$, respectively. Ref.~\cite{Boixo:2014yu} chose the value of $h_{\text{L}}$ to ensure that the minimum ground state gap is lower than the temperature $T=15.5$mK, in order to study a non-trivial interplay between tunneling and thermal activation.\footnote{The problem features one global minimum and one local (false) minimum, which are separated by a tall energy barrier at around the minimum gap point (see Fig.~2 in Ref.~\cite{Boixo:2014yu}); to reach the global minimum from the false minimum, the system state has to transverse the barrier. Such transitions can be modeled as quantum jumps in the quantum jump picture.} This parameter choice means that incoherent effects play a relatively strong role in this problem, which are not well captured by the AME. Thus, similarly to the previous example, the AME is being used here outside of its strict validity domain. We are interested in testing whether it can nevertheless qualitatively capture the correct physical effects. Moreover, direct master equation simulations for such a large system take longer than 24 hours (which is a standard time-window on high-performance clusters), while each quantum trajectory takes less than 24 hours.  We can then exploit many CPU cores to perform many trajectories in parallel. To this end we used $320$ CPU cores and repeated the simulation $16$ times for a total of over $5000$ trajectories.
 
Our simulations (see Fig.~\ref{fig:GoogleGadget}) show how population is lost from the instantaneous ground state to the first excited state near the minimum gap point $s \approx 0.308$.
It also shows a small population revival after the minimum gap is crossed.  As in Sec.~\ref{sec:8qubitchain}, this revival is associated with jumps from the first excited state (or higher states for large $T$) back to the ground state. Encouragingly, despite the perturbative nature of the AME, this revival is qualitatively in agreement with the results of Ref.~\cite{Boixo:2014yu} (see their Fig.~4). The latter work found a stronger revival on the basis of the non-perturbative, non-interacting blip approximation (NIBA), which more accurately captures additional transitions that occur when the energy level broadening is larger than the energy gap between energy levels.

\section{Simulation cost comparison}
\label{sec:timequantitative}

We now provide a cost comparison between the simulations cost of directly solving the AME and the quantum trajectories method. The first two subsections in this section follow Ref.~\cite{Breuer:2002} closely (while adding some details), and we borrow the notation used in that reference. In subsection~\ref{sec:parallel} we provide a new analysis that reveals that the quantum trajectories method can exhibit a scaling advantage ranging between $O(N)$ and $O(N^2)$ over the direct solution of the AME.

 \subsection{Number of trajectories}
The number of trajectories needed can be found from the standard error of the sample mean. As an example, let us consider the standard error $\hat{\sigma}_{t}$ associated with the instantaneous ground state population  $\langle\psi_{0}(t)|\rho(t)|\psi_{0}(t)\rangle$:
\begin{align}
\hat{\sigma}_{t}^2 &= \frac{1}{R(R-1)}\sum_{r=1}^{R}\left(|\langle\psi_{0}(t)|\psi^{r}(t)\rangle|^2  - \hat{M}_t  \right)^2  \label{eq:stderror}\, ,
\end{align}
where $\ket{\psi^r(t)}$ denotes the state associated with trajectory $r$ at time $t$, $\hat{M}_t = \frac{1}{R}\sum_{r=1}^{R}|\langle\psi_{0}(t)|\psi^{r}(t)\rangle|^2$, and $R$ is the total number of trajectories.  By fixing the value of the standard error $\hat{\sigma}_{t}$, the number of necessary trajectories $R$ can then be determined.

\subsection{Cost comparison}
Since we expect $\hat{\sigma}_{t} \sim \frac{1}{\sqrt{R}}$, let us write
\begin{equation}
\label{eq:factor}
\hat{\sigma}_{t}^2 =  \frac{\lambda_{B}(N)}{R} \sim \frac{1}{R}\,,
\end{equation}
where $\lambda_{B}(N) = \frac{1}{R-1}\sum_{r=1}^{R}(\langle\psi^r(t)| B|\psi^r(t)\rangle  - \hat{M}_t)^2$ for an observable $B$ and mean value $\hat{M}_t = \frac{1}{R}\sum_{r=1}^{R}\langle\psi^{r}(t)|B|\psi^{r}(t)\rangle$. The factor $\lambda_{B}(N)$ is a non-increasing function of the system dimension $N$ \cite{Breuer:2002}: 
\begin{equation}
\label{eq:lambdaB}
\lambda_{B}(N) \sim N^{-x} \,,
\end{equation}
where the scaling $x$ depends on the observable: 
\begin{equation}
0 \left(\text{not self-averaging}\right) \leq x \leq 1 \left(\text{strongly self-averaging}\right) \,.
\end{equation}
Thus, to obtain the same standard error for increasing dimension, the number of trajectories need not be increased in general.  This is another advantage of the trajectories method for growing system dimension. Such a phenomenon has also been observed in time-dependent stochastic density functional theory~\cite{gao2015sublinear}.
From Ref.~\cite{Breuer:2002}, the total serial CPU time required for the simulation of the master equation, denoted $T_{\text{AME}}$, versus the stochastic method with $R$ trajectories, denoted $T_{\text{StS}}$, is:
\bes
\label{eq:timecost}
\begin{align}
T_{\text{AME}} &= k_{1} s_{1}(N) N^{\beta} \,, \label{eq:T-AME}\\
T_{\text{StS}} &= k_{2} R(N) s_{2}(N) N^{\alpha} \,,
\label{eq:T-StS}
\end{align}
\ees
where $k_1$ and $k_2$ are constants depending on the specific implementation of each method, $s_{1}(N)$ is the total number of evaluations of $\mathcal{L}_{\textrm{WCL}}[\rho(t)]$ [Eq.~\eqref{eqt:ME2-L}] using the master equation method, and $s_{2}(N)$ is the total number of evaluations of $H_{\text{eff}}(t)\ket{\psi(t)}$ [Eq.~\eqref{eq:Heff}] in a single trajectory.

$R(N)$ in Eq.~\eqref{eq:T-StS} is the minimum number of trajectories needed to obtain a standard error lower than a particular chosen value. To account for the constraint that $R(N)\geq 1$, we rewrite Eq.~\eqref{eq:lambdaB} as $\lambda_B(N) = \Lambda_{B} N^{-x}$, and Eq.~\eqref{eq:factor} as:
\begin{equation}
R(N) =
\begin{dcases} 
      \Bigl\lceil \frac{\Lambda_{B} N^{-x}}{\hat{\sigma}_{t}^2} \Bigr\rceil& \text{\space  for } N < N^{*} \\
      1 & \text{\space  for } N\geq N^{*} 
   \end{dcases}
\end{equation}
where $N^{*} =  \lceil\left({\Lambda_{B}}/{\hat{\sigma}_{t}^2}\right)^\frac{1}{x}\rceil$.   
For $x > 0$, the required number of trajectories decreases with $N$ until $N^{*}$, after which one trajectory gives the expectation value within the desired accuracy.

In general, the number of operations needed to evaluate $\mathcal{L}_{\textrm{WCL}}[\rho(t)]$ relative to the number needed to evaluate $H_{\text{eff}}(t)\ket{\psi(t)}$ differs by a factor of $N$, so that $\beta \approx \alpha + 1$, and Eq.~\eqref{eq:timecost} becomes
\bes
\begin{align}
T_{\text{AME}} &= k_1 s_{1}(N) N^{\alpha + 1} \,,\\
T_{\text{StS}} &=
\begin{dcases} 
      k_2' s_{2}(N) N^{\alpha - x} &\text{\space  for } N < N^{*}  \\
      k_{2} s_{2}(N) N^{\alpha} &\text{\space  for }  N\geq N^{*} 
   \end{dcases}
\end{align}
\ees
where 
\begin{equation}\label{eqt:k2prime}
k_2' = k_2\left(\frac{\Lambda_{B}}{\hat{\sigma}_{t}^2}\right) \,,
\end{equation}
and $k_2'$ hence depends on the required accuracy as well. In many situation $s_{1}(N)$ and $s_{2}(N)$ grow with $N$, but they are roughly equal or grow in same manner with $N$. By dividing these two expressions, we can obtain the ratio of $T_{\text{AME}}/T_{\text{StS}}$, 
\begin{equation}
\frac{T_{\text{AME}}} {T_{\text{StS}}} =
\begin{dcases} 
      \frac{k_1  }{k_2' }N^{1 + x} & \text{\space  for }N < N^{*} \\
      \frac{k_1  }{k_2 }N & \text{\space  for } N\geq N^{*} 
   \end{dcases}  \ . 
\end{equation}
Since $0\leq x \leq 1$, we can write
\begin{equation} 
\label{eqt:timeComparison}
\begin{aligned}
       \frac{k_1}{k_2'}N &\leq \frac{T_{\text{AME}}}{T_{\text{StS}}} \leq  \frac{k_1}{k_2'}N^2 & \text{\space  for } N < N^{*} \,,\\
     &  \frac{T_{\text{AME}}}{T_{\text{StS}}} = \frac{k_1}{k_2}N & \text{\space for } N\geq N^{*} \,. 
\end{aligned} 
\end{equation}

We say that the trajectories method has an advantage over the direct master equation solution if $\frac{T_{\text{AME}}}{T_{\text{StS}}}> 1$. The constant factor $\frac{k_1}{k_2'}$ is typically a small number because it is proportional to the required standard error squared [Eq.~\eqref{eqt:k2prime}]. 
Therefore there is an advantage for the trajectories method when either $N$ is sufficiently large or when a sufficient number of CPU cores $C$ is available (see the next subsection). 
Equation~\eqref{eqt:timeComparison} shows that an advantage
beyond linear in $N$ is attainable for $N<N^\ast$ on a single CPU. The reason is that the number of trajectories needed to achieve a fixed accuracy decreases with increasing system dimension.  For $N>N^\ast$, only one trajectory is required, and the advantage scales as $O(N)$.   

We note that the larger-than-linear advantage only holds if we are interested in estimating operators with the same self-averaging property. This is in contrast to evolving the entire density matrix, as in the AME, which allows the expectation value of any observable to be calculated.  If we demand this same capability from the trajectories approach, then only the linear advantage holds.
 
\subsection{Parallel implementation}
\label{sec:parallel}

The stochastic wave function method is very well-suited for parallel computing implementations. The communication needed between each core is minimal since each trajectory is simulated independently. Assuming $C$ CPU cores are used, where $C \leq R(N)$, we can adjust the time cost [Eq.~\eqref{eq:T-StS}] for the stochastic method to
\begin{equation}
T_{\text{StS}} = k_{2} \frac{R(N)}{C} s_{2}(N) N^{\alpha} \,.
\end{equation}
Note that the number of cores $C$ is held constant, i.e., is independent of the system dimension $N$. Therefore, we can update Eq.~\eqref{eqt:timeComparison} to:
\begin{equation}
\begin{aligned}
       \frac{k_1}{k_2'}C N &\leq \frac{T_{\text{AME}}}{T_{\text{StS}}} \leq  \frac{k_1}{k_2'}C N^2 & \text{\space  for } N < N^{\star} \,,\\
     &  \frac{T_{\text{AME}}}{T_{\text{StS}}} = \frac{k_1}{k_2}  N & \text{\space for } N\geq N^{\star} \,.
      \end{aligned} 
\end{equation}
where $N^{\star} =  \lceil\left({\Lambda_{B}}/\left({C \hat{\sigma}_{t}^2} \right) \right)^\frac{1}{x}\rceil$.  Here $N^{\star}$ is the system dimension where $R(N^{\star}) = C$, and one execution of the $C$ parallel CPU cores is enough to obtain the desired standard error.  

Again, the larger-than-linear advantage in $N$ only holds if we are interested in estimating operators with the same self-averaging property, and otherwise we can only expect a linear advantage in $N$.

\section{Conclusions and Outlook}
\label{sec:VI}
In this work, we have shown how quantum trajectories (in the form of quantum jumps) can be unravelled from the adiabatic master equation. We have described and demonstrated a simulation procedure in terms of the waiting time distribution that reproduces the results of the master equation for examples involving $8$ and $16$-qubit systems. 
Direct master equation simulations for the $16$-qubit example would take a long time, but the simulation of the quantum trajectories remains computationally feasible for larger system dimensions by allowing us to simulate many trajectories in parallel. 
A scaling cost comparison of the two methods shows that, generically, the quantum trajectories method yields an improvement by a factor linear in the system dimension $N$ over directly solving the adiabatic master equation. However, the trajectories method can be expected to be up to a factor $c N^2$ faster than a direct simulation of the master equation if only the expectation value of specific self-averaging observables is desired. Here $c$ is a constant proportional to the number of parallel processes and the target standard error.
 
We therefore believe this approach will be particularly useful in enabling the study of larger systems than has been possible using a direct simulation of the AME. 

In addition, the quantum trajectories method offers fresh physical insight into the nature of individual trajectories and their statistics, which may become a helpful tool in interpreting computational bottlenecks in quantum annealing and adiabatic quantum computing.

Finally, while we did not address this in the present work, the quantum trajectories approach is well known to be a convenient path towards continuous measurement and the inclusion of quantum feedback control~\cite{Wiseman:book}. This approach might in the future provide a path towards error correction of adiabatic quantum computing, e.g., by formulating control targets that push the system back to the ground state after diabatic or thermal transitions.

\acknowledgements

The research is based upon work (partially) supported by the Office of
the Director of National Intelligence (ODNI), Intelligence Advanced
Research Projects Activity (IARPA), via the U.S. Army Research Office
contract W911NF-17-C-0050. The views and conclusions contained herein are
those of the authors and should not be interpreted as necessarily
representing the official policies or endorsements, either expressed or
implied, of the ODNI, IARPA, or the U.S. Government. The U.S. Government
is authorized to reproduce and distribute reprints for Governmental
purposes notwithstanding any copyright annotation thereon. Computation for the work described in this paper was supported by the University of Southern California's Center for High-Performance Computing.


\appendix

\section{Error estimates}
\label{app:A}

Let us assume that the system is in a pure state at time $t$, i.e., $\rho_S(t) = \ketbra{\psi(t)}$, and let us consider a single time-step. In a single trajectory, the evolution of $\ket{\psi(t)}$ involves two possibilities: no jump or a jump. The ensemble average of trajectories after one finite time-step mainly involves two kinds of errors: the error associated with the norm of the state vector (Sec.~\ref{sssec:en}), and the error associated with the probability elements in a finite time-step (Sec.~\ref{sssec:ep}).

\subsection{Error associated with the norm squared of a no-jump trajectory}
\label{sssec:en}
The Schr\"{o}dinger equation of the effective Hamiltonian in the case of a no-jump trajectory is given by
\begin{equation}
\label{eqt:Schrapp2}
\frac{d\ket{\psi(t)}}{dt} = -i H_{\text{eff}}(t) \ket{\psi(t)} \ .
\end{equation}
The resulting state after one time-step $\Delta t$ is:
\begin{equation}
\label{eqt:euler} 
\ket{\tilde{\psi}(t+\Delta t)} =V_{\text{eff}}(t+\Delta t, t)\ket{\psi(t)}  \ ,
\end{equation}
where 
\beq 
V_{\text{eff}}(t+\Delta t, t) = {\cal T}\exp\left[-i\int_t^{t+\Delta t}H_{\text{eff}}(t')dt'\right]
\eeq 
is a non-unitary contractive evolution operator ($\vertiii{ V_{\text{eff}}(t+\Delta t, t)} \leq 1$ for every induced norm) and $\ket{\tilde{\psi}(t+\Delta t)}$ is unnormalized since $H_{\text{eff}}(t)$ is not hermitian.
We can expand the time-ordered exponential for $V_{\text{eff}}(t+\Delta t, t)$ as
\onecolumngrid
\begin{align}
\label{Eqt: TDtaylor}
&{}{\cal T}\exp\left[-i\int_t^{t+\Delta t}H_{\text{eff}}(t')dt'\right]    \\
&=\mathds{1} - i \int_{t}^{t+\Delta t} dt' H_{\text{eff}}(t')   + \left(-i\right)^2 \frac{1}{2!} \mathcal{T}\left(\int_{t}^{t+\Delta t} dt' H_{\text{eff}}(t')\right)^2 +
\left(-i\right)^3 \frac{1}{3!}\mathcal{T}\left(\left(\int_{t}^{t+\Delta t} dt' H_{\text{eff}}(t')\right)^3\right) \cdots \nonumber
\end{align}
For any $H_{\text{eff}}(t)$ that is $C^{K}$ on an open interval containing $[0, t_f]$ ($H^{(k)}_{\text{eff}}(t)$ is continuous and bounded for $1\leq k \leq K$):
\begin{equation}
\int_{t}^{t+\Delta t} dt' H_{\text{eff}}(t') = 
\sum\limits_{k = 0}^{K-1}\frac{H^{(k)}_{\text{eff}}(t)}{(k+1)!}(\Delta t)^{k+1} + O((\Delta t)^{K+1})\, ,
\end{equation}
as $\Delta t \rightarrow 0$. Assume $H_{\text{eff}}(t)$ is $C^{K}$ and $K \geq 2$. Let $\dot{H}_{\eff}(t)\equiv H^{(1)}_{\text{eff}}(t)$ denote the time derivative of $H_\eff(t)$.
We have:
\bes
\begin{align}
\label{Eqt: integral}
\int_{t}^{t+\Delta t} dt' H_{\text{eff}}(t')&= H_{\text{eff}}(t)\Delta t + \frac{1}{2!}\dot{H}_{\text{eff}}(t) (\Delta t)^2 + O((\Delta t)^3) \,,\\
\mathcal{T}\left(\int_{t}^{t+\Delta t} dt' H_{\text{eff}}(t')\right)^2 &= 2\int_{t}^{t+\Delta t} dt' H_{\text{eff}}(t')  \int_{t}^{t'} dt'' H_{\text{eff}}(t'') \nonumber\\
&= H_{\text{eff}}(t)^2(\Delta t)^2 + \frac{1}{2}H_{\text{eff}}(t)\dot{H}_{\text{eff}}(t)(\Delta t)^3 + \frac{1}{2}\dot{H}_{\text{eff}}(t)H_{\text{eff}}(t)(\Delta t)^3 + O((\Delta t)^4)\ \,, \\
\mathcal{T}\left(\int_{t}^{t+\Delta t} dt' H_{\text{eff}}(t')\right)^3 &= H_{\text{eff}}(t)^3(\Delta t)^3 + O((\Delta t)^4) \,.
\end{align}
\ees
Therefore, we have, to second order in $\Delta t$:
\beq
\label{eq:A7}
V_{\text{eff}}(t+\Delta t, t) = \mathds{1} -i \left( H_{\text{eff}}(t)\Delta t + \frac{1}{2!}\dot{H}_{\text{eff}}(t) (\Delta t)^2 \right) + \left(-i\right)^2 \frac{1}{2!} H_{\text{eff}}(t)^2(\Delta t)^2  + O((\Delta t)^3)\ ,
\eeq
so that:
\begin{align}
&{} V^{\dagger}_{\text{eff}}(t+\Delta t, t) V_{\text{eff}}(t+\Delta t, t) \nonumber \\
=&{}  \mathds{1} + i\Delta t \left(H^{\dagger}_{\text{eff}}(t) - H_{\text{eff}}(t)\right) 
+ \frac{1}{2!}(\Delta t)^2\left[ i\dot{H}^{\dagger}_{\text{eff}}(t) - i\dot{H}_{\text{eff}}(t) 
- \left(H^{\dagger 2}_{\text{eff}}(t) + H^2_{\text{eff}}(t)\right)\right] + O((\Delta t)^3) \ . 
\end{align}
We approximate $N(t,\Delta t)\equiv \|\ket{\tilde{\psi}(t+\Delta t)} \|^2$ by:
\beq
\label{Eqt: Step2}
N(t,\Delta t) = \braket{\tilde{\psi}(t+\Delta t)|\tilde{\psi}(t+\Delta t)} 
\approx \bra{\psi(t)}\left(\mathds{1} + i \Delta t\left(H^{\dagger}_{\text{eff}}(t) - H_{\text{eff}}(t) \right)\right)\ket{\psi(t)} \equiv E(t,\Delta t)\, .
\eeq
The approximation error is given by:
\bes
\begin{align}
\delta (t,\Delta t) &\equiv N(t,\Delta t)-E(t,\Delta t)\\
 &= \frac{1}{2}(\Delta t)^2\bra{\psi(t)}\left[ i\dot{H}^{\dagger}_{\text{eff}}(t) - i\dot{H}_{\text{eff}}(t) 
- \left(H^{\dagger 2}_{\text{eff}}(t) + H^2_{\text{eff}}(t)\right)\right] \ket{\psi(t)}+ O((\Delta t)^3) \\
&\leq \frac{(\Delta t)^2}{2}\vertiii{i(\dot{H}^{\dagger}_{\text{eff}}(t) - \dot{H}_{\text{eff}}(t)) - ( H^{\dagger 2}_{\text{eff}}(t) + H^2_{\text{eff}}(t))} + \vertiii{O((\Delta t)^3)}\ ,
\label{Eqt: errornorm}
\end{align}
\ees
\twocolumngrid
where $\|\cdot\|$ is the operator norm (largest singular value).

Eq.~\eqref{Eqt: errornorm} gives the relation between the error of the norm square approximation and the time-step. It is also helpful to consider the sources of error here as they will be used later. This error mainly arises from two sources during the truncations of Taylor expansion: 
\begin{enumerate}
\item The truncation of the Taylor expansion of the integral [Eq.~\eqref{Eqt: integral}] to keep only $H_{\text{eff}}(t)\Delta t$. This turns Eq.~\eqref{Eqt: TDtaylor} into $\exp\left(-iH_{\text{eff}}(t)\Delta t\right)$.  For this to hold, we require 
\begin{align}
\label{Eqt: case1}
\left\lVert \frac{1}{2}\dot{H}_{\text{eff}}(t) (\Delta t)^2 
+ O((\Delta t)^3) \right\rVert \ll  \lVert H_{\text{eff}}(t)\Delta t \rVert \, ,
\end{align}
implying:
\begin{align}
\Delta t \ll 2\frac{\vertiii{H_{\text{eff}}(t)}}{\vertiii{\dot{H}_{\text{eff}}(t)}}  \, ,
\label{Eqt: 1stcon}
\end{align}
assuming $\dot{H}_{\text{eff}}(t)\neq 0$; if it is then the condition becomes $\Delta t \ll \left(\frac{K!\vertiii{H_{\text{eff}}(t)}}{\vertiii{{H^{(K)}_{\text{eff}}(t)}}}\right)^{1/K}$ for the lowest value of $K$ such that $H^{(K)}_{\text{eff}}(t)\neq 0$, where the superscript denotes the $K$-th derivative.

\item Keeping only the first order term in Eq.~\eqref{eq:A7} afterwards, i.e., $\mathds{1} - iH_{\text{eff}}(t)$.  
This requires, in addition to Eq.~\eqref{Eqt: 1stcon}:
\begin{align}
\label{Eqt: case2}
\left\lVert\frac{1}{2}H^2_{\text{eff}}(t)(\Delta t)^2 + O((\Delta t)^3)
\right\rVert \ll  \vertiii{H_{\text{eff}}(t)\Delta t} \, ,
\end{align}
\end{enumerate}
implying, for all $t$ such that the denominators do not vanish:
 \beq 
 \label{eqt:Deltat2}
 \Delta t \ll  \min \left\{ \frac{\vertiii{H_{\text{eff}}(t)}}{\vertiii{\dot{H}_{\text{eff}}(t)}}, \frac{1}{\vertiii{H_{\text{eff}}(t)}} \right\} \ ,
 \eeq
where we ignored the factor of $2$.
In conclusion, the norm after one time-step is related to the jump probabilities as:
\bes
\begin{align}
N(t,\Delta t) &= 
 \bra{\psi(t)}\left(\mathds{1} + i (\Delta t)\left(H^{\dagger}_{\text{eff}}(t) - H_{\text{eff}} \right)\right)\ket{\psi(t)} \notag \\
 &\qquad + \delta(t,\Delta t) \\
&= 1 - \Delta t \sum\limits_{i}\braket{{A}_{i}^{\dagger}(t) {A}_{i} (t)} + \delta(t,\Delta t) \\
&= 1 - \Delta p(t) + \delta(t,\Delta t) \,,\label{Eqt: squareofnorm}
\end{align}
\ees
where we used Eq.~\eqref{eq:Heff} and defined the (approximate) jump probabilities as:
 \bes
 \label{Eqt: typeofprob}
\begin{align}
&{}\Delta p(t)  = \sum_i \Delta p_i(t) \\
&{}\Delta p_i(t) = \Delta t \braket{ {A}_{i}^{\dagger}(t) {A}_{i} (t)}  \, .
\label{eqt:typeofprob}
\end{align}
\ees
This explains Eqs.~\eqref{eqt:squareofnorm} and~\eqref{eq:jumprate}.

\subsection{Error associated with probability elements}
\label{sssec:ep}
We have defined the $\Delta p_i(t)$ and $\Delta p(t)$ in a fixed time-step in Eq.~\eqref{Eqt: typeofprob}.
Note that even in the time-independent case, where both the Hamiltonian and Lindblad operators are time-independent, $1- \Delta p(t)$ is never exactly equal to the norm squared of the state vector after one time-step, and $\Delta p_i(t)$ is not exactly the jump probability inside the time-step. They are only approximations. 

For a finite time-step $\Delta t$, with $p_0$ the probability of having no jump inside the interval $[t, t+\Delta t]$ and $p_1$ the probability of having one jump inside the interval $[t, t+\Delta t]$, we have:
\beq
p_{0} = e^{-\int_t^{t+\Delta t} \lambda(t') dt'} = 1-p_{1}  \,.
\label{eq:p_0}
\eeq
Note that as $\Delta t \rightarrow 0$, $p_{0} = 1 - \lambda(t)\Delta t + o(\Delta t)$, $p_{1} \simeq \lambda(t)\Delta t$.
We shall focus on the case where the time-step is sufficiently small such that the probability of two or more jumps occurring within a single time-step is negligible.\footnote{E.g., the quantum jump can be described by a Poisson process with a state-dependent inhomogeneous jump rate, with two or more jumps as successive one and no-jump processes, and $\sum_{n\geq 2}p_{n} = o(\Delta t)$ as $\Delta t\rightarrow 0$.} 

First, we can expand the exponential as
\begin{align}
p_0 
&=1 - \int_t^{t+\Delta t} 
\lambda(t')dt' + \frac{1}{2}\left(\int_t^{t+\Delta t} \lambda(t') dt'
\right)^2 \cdots \nonumber
\end{align}
During the time window from $t$ to $t+\Delta t$, $\lvert\psi(t')\rangle$ [required to calculate $\lambda(t')$] is obtained by solving the Schr\"{o}dinger equation with the effective Hamiltonians [Eq.~\eqref{eqt:Schrapp2}] and renormalizing the solution during the integration. As we show in Sec.~\ref{sec:C}, in any finite interval the norm squared of the unnormalized state vector [Eq.~\eqref{Eqt: Step2}] is equal to $p_0$. This is the reason why we can use the waiting time distribution (or Gillespie algorithm~\cite{gillespie1977exact}) as our simulation method in the main text.

Second, since
\begin{align}
\int_{t}^{t+\Delta t} dt' \lambda(t') = \sum\limits_{k = 0}^{\infty}\frac{\lambda^{(k)}(t)}{(k+1)!}(\Delta t)^{k+1} \ ,
\end{align}
we have
\begin{align}
\label{eqt:p0allterms}
= 1 - \lambda(t)\Delta t + e_p \ ,
\end{align} 
where the error $e_p$ associated with the probability elements in a fixed time-step is:
\begin{equation}
e_p = - \frac{1}{2}\frac{d}{dt}
\lambda(t)(\Delta t)^2 + \frac{1}{2}
\lambda^2(t)(\Delta t)^2 + O((\Delta t)^3) \,.
\end{equation}
This should be much smaller than the first order term $\lambda(t) \Delta t$. Therefore, we need 
\begin{align}
\Delta t \ll \left|\frac{ \lambda(t) }{  \lambda^2(t)-\frac{d}{dt} \lambda(t)}\right|  \, .
\label{Eqt: Poissoncon}
\end{align}
In the time-independent case, this reduces to 
\beq
\Delta t \ll \frac{1}{ \lambda} \ .
\eeq

\section{Proof of equivalence between the master equation and trajectories formulations}
\label{app:B}

Our goal in this section is to show how the master equation, Eq.~\eqref{eqt:effdiagonalform}, can be recovered from the quantum trajectories formulation, and to find a bound on the time-step $\Delta t$. This generalizes the proof for the time-independent case found in~\cite{molmer1993monte}.

\subsection{To jump or not to jump}
\label{sssec:2}
The probability elements $\Delta p(t)$ and $\Delta p_i (t)$ are important for determining whether a jump occurs and if a jump does occur, which jump type occurs. 
In order to determine if a jump occurs or not, we draw a random number $\epsilon$, uniformly distributed between 0 and 1. If $\Delta p(t) < \epsilon$, which is almost always the case since $\Delta p(t)$ is very small, no jump occurs. In the case of no jump, $\ket{\psi(t)}$  evolves according to the effective Schr\"{o}dinger equation, Eq.~\eqref{eqt:Schrapp2}. At time $t+\Delta t$ we simply renormalize the solution of Eq.~\eqref{eqt:euler}:
\bes
\begin{align}
\left|\psi(t + \Delta t)\right> &= \frac{1}{\sqrt{\braket{\tilde{\psi}(t+\Delta t)| \tilde{\psi}(t+\Delta t)}}}\ket{\tilde{\psi}(t+\Delta t)}\\&\stackrel{\eqref{Eqt: squareofnorm}}{=} \frac{1}{\sqrt{1 - \Delta p(t) + \delta}}\ket{\tilde{\psi}(t+\Delta t)} \ .
\end{align}
\ees

If $\Delta p(t) > \epsilon$, the state undergoes an abrupt jump and we choose the new wavefunction among the different states $A_{i}\ket{\psi(t)}$ and renormalize:
\bes
\begin{align}
\left|\psi(t + \Delta t)\right> &= \frac{{A}_{i}(t)\ket{\psi(t)}}{\sqrt{\bra{\psi(t)} {A}_{i}^{\dagger}(t) {A}_{i} (t)\ket{\psi(t)}}} \\
&\stackrel{\eqref{eqt:typeofprob}}{=} 
\sqrt{\frac{\Delta t}{\Delta p_i(t)}} {A}_{i}(t)\ket{\psi(t)}\ .
\end{align}
\ees
Which type of jumps occurs is determined according to the probability 
\bes
\begin{align}
\Pi_i(t) &= \frac{\Delta p_i(t)}{\Delta p(t)} = \frac{\braket{\psi(t)|{A}_{i}^{\dagger}(t){A}_{i}(t)|\psi(t)}\Delta t}{\sum\limits_{i}\braket{\psi(t)|{A}_{i}^{\dagger}(t){A}_{i}(t)|\psi(t)}\Delta t} \\
&= \frac{\braket{{A}_{i}^{\dagger}(t){A}_{i}(t)}}{\lambda(t)}\ ,
\label{eq:Pi(t)}
\end{align}
\ees
where 
\beq
\lambda(t) = \sum_{i}\langle {A}_{i}^{\dagger}(t){A}_{i}(t)\rangle 
\label{eq:lambda}
\eeq
is the time-dependent jump rate.

\subsection{Averaging over trajectories}
\label{sssec:3}

Let $H_{\text{eff}}(t)$ be $C^{K}$ with $K \geq 2$. 
We first express the mean value $\bar{\sigma}_{S}(t)$ as a sum over the non-Hermitian evolution [with probability $1 - \Delta p(t)$] and the jump trajectories [with probability $\Delta p(t)$], so that as $\Delta t \rightarrow 0$ we have: 
\onecolumngrid
\bes
\begin{align}
\label{Eqt: averaging}
\bar{\sigma}_{S}(t + \Delta t) 
&= \left[1 - \Delta p(t)\right]\frac{\ket{\tilde{\psi}(t + \Delta t)}}{\sqrt{1 - \Delta p(t) + \delta}}\frac{\bra{\tilde{\psi}(t + \Delta t)}}{\sqrt{1 - \Delta p(t) + \delta}}  \nonumber \\
& + \Delta p(t) \sum\limits_{i} \Pi_{i}(t) \sqrt{\frac{\Delta t}{\Delta p_i(t)}}{A}_{i}(t)  \ket{\psi(t)}\sqrt{\frac{\Delta t}{\Delta p_i(t)}}\bra{\psi(t)}{A}_{i}^{\dagger}(t) \\
&\stackrel{\eqref{eq:Pi(t)}}{=}   \frac{1 - \Delta p(t)}{1 - \Delta p(t) + \delta}\ket{\tilde{\psi}(t + \Delta t)}\bra{\tilde{\psi}(t + \Delta t)}
+ \Delta t \sum\limits_{i}{A}_{i}(t)\sigma_{S}(t){A}_{i}^{\dagger}(t) 
\label{Eqt: noninnerproductform}
\end{align}
\ees
Combining Eq.~\eqref{eqt:euler} with Eq.~\eqref{eq:A7} we have:
\beq
\ket{\tilde{\psi}(t+\Delta t)} =
\left[ \mathds{1} -i \left( H_{\text{eff}}(t)\Delta t + \frac{1}{2!}\dot{H}_{\text{eff}}(t) (\Delta t)^2 \right) + \left(-i\right)^2 \frac{1}{2!} H_{\text{eff}}(t)^2(\Delta t)^2  + O((\Delta t)^3) \right] \ket{\psi(t)}  \ .
\eeq
Recall that in Eq.~\eqref{eqt:Deltat2} we gave conditions allowing us to neglect the $O((\Delta t)^2)$ terms.
Thus: 
\begin{align}
\label{eq:B3}
\bar{\sigma}_{S}(t + \Delta t) 
&=  \frac{1 - \Delta p(t)}{1 - \Delta p(t) + \delta}\left( \mathds{1} - i H_{\text{eff}}(t) \Delta t 
+ O((\Delta t)^2)\right)\bar{\sigma}_{S}(t) \left(\mathds{1} + i H^{\dagger}_{\text{eff}}(t) \Delta t
+O((\Delta t)^2)\right)\\
&\phantom{==}+ \Delta t \sum\limits_{i}{A}_{i}(t)\bar{\sigma}_{S}(t){A}_{i}^{\dagger}(t)\nonumber  \, .
\end{align} 
\begingroup
\allowdisplaybreaks
where we have replaced $\sigma_{S} (t)$ by $\bar{\sigma}_{S} (t)$ after averaging over many trajectories. 
Rearranging this expression into a form that exposes the terms that will become the master equation,
the expression for the averaged state at $t + \Delta t$ becomes:
\bes
\begin{align}
\bar{\sigma}_{S}(t + \Delta t) &= \bar{\sigma}_{S}(t) + i\Delta t\left(\bar{\sigma}_{S}(t)H_{\text{eff}}^{\dagger}(t) - H_{\text{eff}}(t)\bar{\sigma}_{S}(t) \right)+ \Delta t \sum_{i} {A}_{i}(t) \bar{\sigma}_{S}(t){A}_{i}^{\dagger}(t) \\
&\phantom{==}
- \frac{\delta}{1 - \Delta p(t) + \delta} \left[ \bar{\sigma}_{S}(t) +  i\bar{\sigma}_{S}(t)\left(H_{\text{eff}}^{\dagger}(t) -H_{\text{eff}}(t)\right) \bar{\sigma}_{S}(t) \Delta t \right] + O((\Delta t)^2)\ .
\label{eq:B5b}
\end{align}
\ees
\twocolumngrid
Note that $\delta$ is $O((\Delta t)^2)$ as $\Delta t \rightarrow 0$ [Eq.~\eqref{Eqt: errornorm}], and $\Delta p(t) = \Delta t \sum_{i}\braket{\psi(t)|{A}_{i}^{\dagger}(t){A}_{i}(t)|\psi(t)}$ is $O(\Delta t)$ as $\Delta t \rightarrow 0$, so that:
\bes
\begin{align}
\frac{\delta}{1 - \Delta p(t) + \delta} &= 
\frac{O((\Delta t)^2)}{1 - O(\Delta t) + O((\Delta t)^2)} \\
&=O((\Delta t)^2)\, . 
\end{align} 
\ees
Therefore line~\eqref{eq:B5b} can be absorbed into $O((\Delta t)^2)$, and we are left with:
\begin{align}
\frac{\bar{\sigma}_{S}(t + \Delta t) - \bar{\sigma}_{S}(t)}{\Delta t} =&
-i\left(H_{\text{eff}}(t)\bar{\sigma}_{S}(t) - \bar{\sigma}_{S}(t)H_{\text{eff}}^{\dagger}(t) \right)\notag \\
&+  \sum_{i} {A}_{i}(t) \bar{\sigma}_{S}(t){A}_{i}^{\dagger}(t) 
+ O(\Delta t) \ ,
\end{align}
which becomes the master equation, Eq.~\eqref{eqt:effdiagonalform}, in the $\Delta t \to 0$ limit.

\subsection{Upper bound on $\Delta t$}
\label{ssec:upperbound}
The above proof takes $\Delta t \rightarrow 0$. We would like to know how small the time-step $\Delta t$ should be in order for the approximations made to be valid. In Eq.~\eqref{eq:B3}, we expanded the time-ordered exponential, and kept only the first order terms. This is equivalent to the criteria in Eqs.~\eqref{Eqt: case1} and ~\eqref{Eqt: case2}, summarized as a single condition in Eq.~\eqref{eqt:Deltat2}. 
As shown in Sec.~\ref{sssec:en}, this also automatically makes the error in the norm squared approximation $\delta$ small. We also need to satisfy Eq.~\eqref{Eqt: Poissoncon}, in order to accurately approximate the probability elements. Taken together, therefore:
 \beq 
 \Delta t \ll  \min \left\{ \frac{\vertiii{H_{\text{eff}}(t)}}{\vertiii{\dot{H}_{\text{eff}}(t)}}, \frac{1}{\vertiii{H_{\text{eff}}(t)}}, \left|\frac{ \lambda(t) }{  \lambda^2(t)-\dot \lambda(t)}\right| \right\} \ .
 \label{eq:B7}
 \eeq
In practice, choosing a constant time-step that satisfies Eq.~\eqref{eq:B7} in the whole timespan $[0, t_f]$ is sufficient, though one might prefer to implement an adaptive time-step tailored to the instantaneous value of the R.H.S. 
\section{On the validity of waiting times (quantum time-dependent operators)}
\label{sec:C}
Here we show the validity of using the waiting time distribution in the case of time-dependent operators. The argument presented here is based on Ref.~\cite{Breuer:2002} and we extend it to the time-dependent case.

Let us denote by $\ket{\psi(t)}$ and $\ket{\tilde{\psi}(t)}$ the normalized and unnormalized state vectors respectively, and let us assume they are equal at time $t$. This can happen when $t = 0$ or any time immediately after each jump. Let $t^+ \equiv t+\tau$, where $\tau$ can be as large as is possible until the next jump occurs,  and
\begin{equation}
V_{\text{eff}}(t^+, t) = {\cal T}\exp\left[-i\int_t^{t^+}H_{\text{eff}}(t')dt'\right] \ ,
\end{equation}
Then:
\bes
\begin{align}
\ket{\tilde{\psi}(t + \tau)} &=  
 V_{\text{eff}}(t^+, t)\ket{\psi(t)}  \ ,\\
\ket{\psi(t^+)} &=  
 \frac{V_{\text{eff}}(t^+, t)\ket{\psi(t)}}{\left\lVert V_{\text{eff}}(t^+, t)\ket{\psi(t)} \right\rVert} \,.
 \label{eq:C2b}
\end{align}
\ees
Then, starting from $t$, for any future $t^{+} > t$, we have
\onecolumngrid
\bes
\begin{align}
\frac{d\phantom{,}}{dt^{+}}\left\lVert \ket{\tilde{\psi}(t^{+})} \right\rVert^2 &= \phantom{}\frac{d\phantom{,}}{dt^{+}}\left\lVert{\cal T}\exp\left[-i\int_t^{t^+}H_{\text{eff}}(t')dt'\right]\ket{\psi(t)}\right\rVert^2 \\
&= \frac{d\phantom{,}}{dt^{+}}\bra{\psi(t)}V_\eff^{\dagger}(t^+,t) V_\eff(t^+,t)\ket{\psi(t)} \\
&=\bra{\psi(t)}V_\eff^{\dagger}(t^+,t)(+i)H^{\dagger}_{\text{eff}}(t^{+})V_\eff^{}(t^+,t)\ket{\psi(t)} +
\bra{\psi(t)}V_\eff^{\dagger}(t^+,t)(-i)H_{\text{eff}}(t^{+})V_\eff^{}(t^+,t)\ket{\psi(t)} \\
&=-\bra{\psi(t)}V_\eff^{\dagger}(t^+,t)\left(\sum\limits_{i}{A}^{\dagger}_{i}(t^{+}){A}_{i}(t^{+})\right)V_\eff^{}(t^+,t)\ket{\psi(t)} \\
&=  -\left\lVert V_{\text{eff}}(t^{+}, t)\ket{\psi(t)} \right\rVert^2 \sum\limits_{i} \bra{\psi(t^{+})} {A}^{\dagger}_{i}(t^{+}){A}_{i}(t^{+}) \ket{\psi(t^{+})}  \, ,
\end{align}
\ees
\twocolumngrid
where in the last equality we used Eq.~\eqref{eq:C2b}.

Let $N(t^{+}) \equiv \| \ket{\tilde{\psi}(t^{+})} \|^2$, as in Eq.~\eqref{Eqt: Step2}.  We have
\begin{equation}
\frac{d\phantom{,}}{dt^{+}} N(t^{+}) = -N(t^{+})  \lambda(t^+)   \, ,
\end{equation}
where $\lambda$ is the time-dependent jump rate [Eq.~\eqref{eq:lambda}].
The solution to this differential equation with the initial condition $N(t) = 1$ is
\begin{equation}
N(t^{+}) = \exp\left(-\int_t^{t^{+}} \lambda(t') dt'\right) = p_0(t^{+})  \, ,
\end{equation}
where $p_0$ is the probability of not having any jump inside the interval [Eq.~\eqref{eq:p_0}], which we have now shown to be equal to the the norm squared of the unnormalized state vector for any finite interval $[t, t+\tau]$. 

No commutators of operators at different times appears in the derivation. The use of the waiting time distribution is therefore valid for time-dependent operators as long as the correlation matrix is positive.

\section{Effect of the annealing schedule on the $8$-qubit chain example} 
\label{app:Chain}

We provide the functional form for the (D-Wave 1) annealing schedule functions $A(t)$ and $B(t)$ used in Sec.~\ref{sec:8qubitchain} in Fig.~\ref{fig:5a}.  We compare this non-linear schedule to a linear annealing schedule with an overall energy scale chosen to closely match the energy scale at which the first annealing schedule curves cross.  The dynamics associated with the linear annealing schedule [shown in Fig.~\ref{fig:5b}] is somewhat different than the non-linear one. This can be attributed to the differences between the two schedules. In order to ensure that the two sets of schedules intersect at the same energy scale, this requires the linear schedule to start from a significantly smaller energy scale.  This results in some important effects on the dynamics.  First, this energy scale is not large enough to ensure that the thermal state at $s = 0$ has negligible weight on excited states.  Since we use the ground state as the initial state of the simulation, which is now sufficiently different from the thermal state at $s=0$, the dissipative dynamics causes visible changes in the state immediately, which can be see as both the dip in Fig~\ref{fig:5a} near $s=0$ and the large number of excitations at the beginning of the anneal in Fig.~\ref{fig:5c}.  Second, even after this initial dip, the lower energy scale associated with the linear schedule means that thermal depopulation of the ground state occurs generally sooner relative to the non-linear schedule studied in the main text, although at a lower rate because of the linear form of the schedule.  Furthermore, because the transverse field remains significant compared to the Ising term for longer in the anneal than in the linear case, repopulation of the ground state due to thermal relaxation occurs for a longer period of time.  However, ultimately, the ground state probability at the end of the anneal is not significantly different than that shown in Fig.~\ref{fig:3qubits8qubits}. The jump statistics are shown in Fig.~\ref{fig:5c}, and closely resemble the non-linear case shown in the inset of Fig.~\ref{fig:8qubitssingle}.

\begin{figure*}[htb!] 
\centering
\subfigure[\ ]{\includegraphics[width=0.32\textwidth]{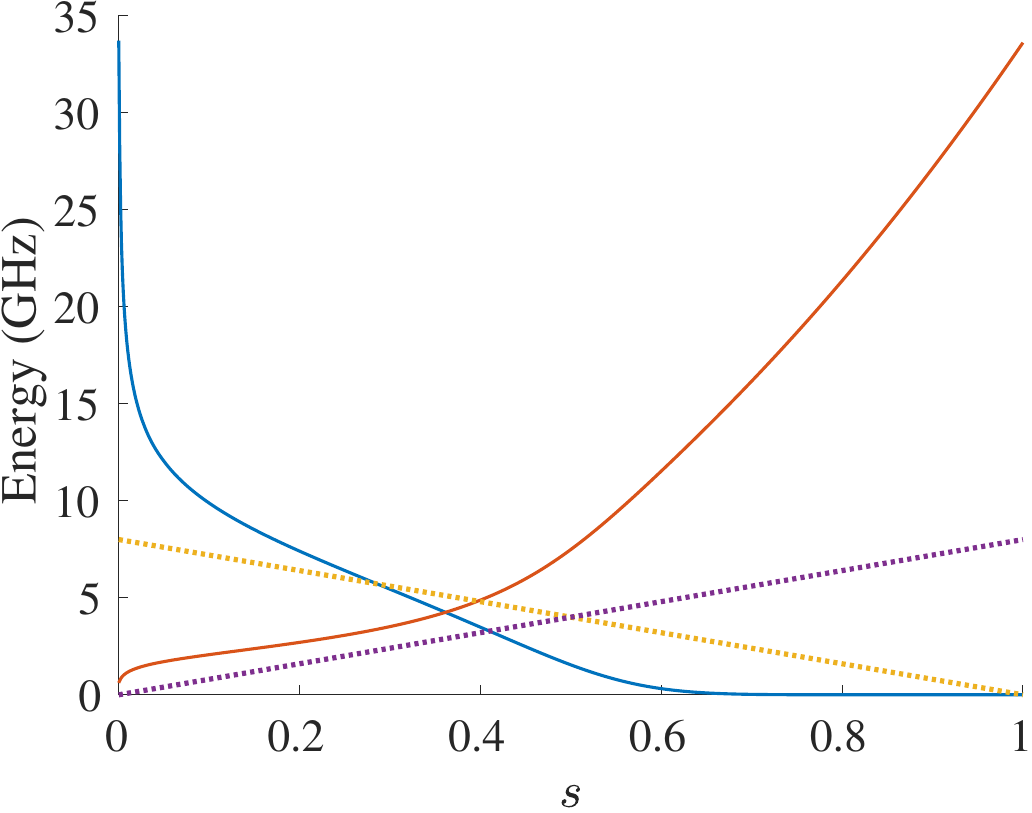}\label{fig:5a}} 
\subfigure[\ ]{\includegraphics[width=0.32\textwidth]{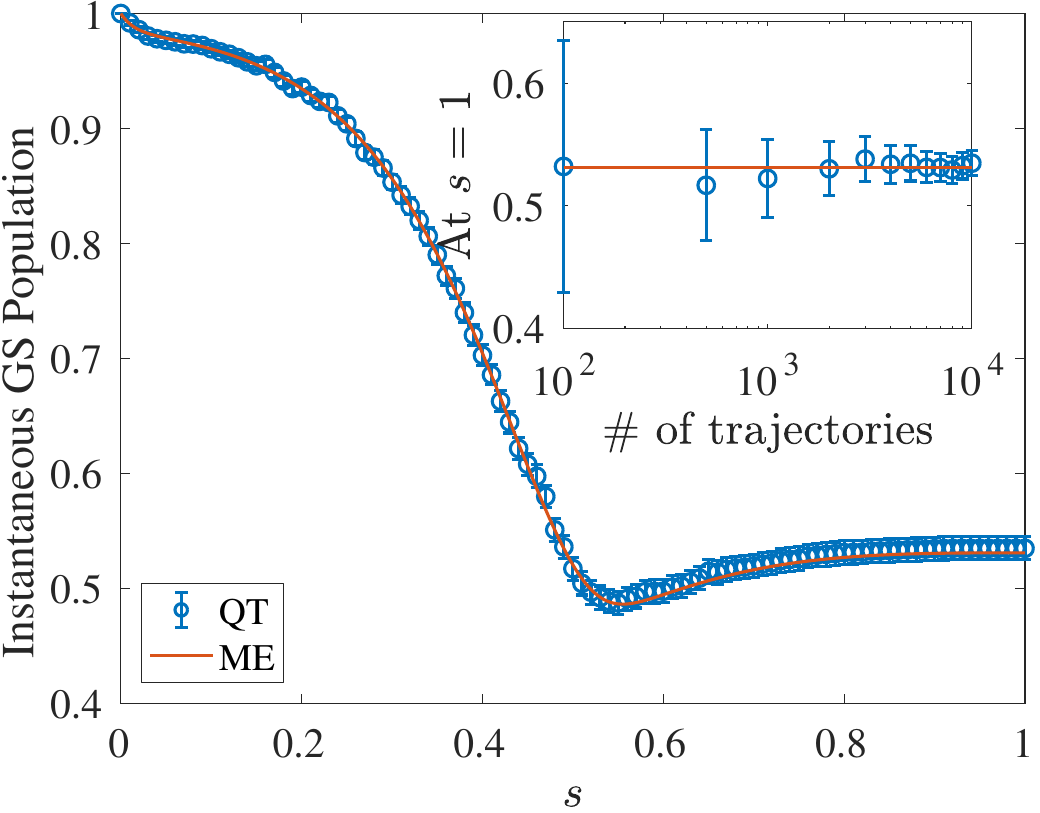}\label{fig:5b}} 
\subfigure[\ ]{\includegraphics[width=0.32\textwidth]{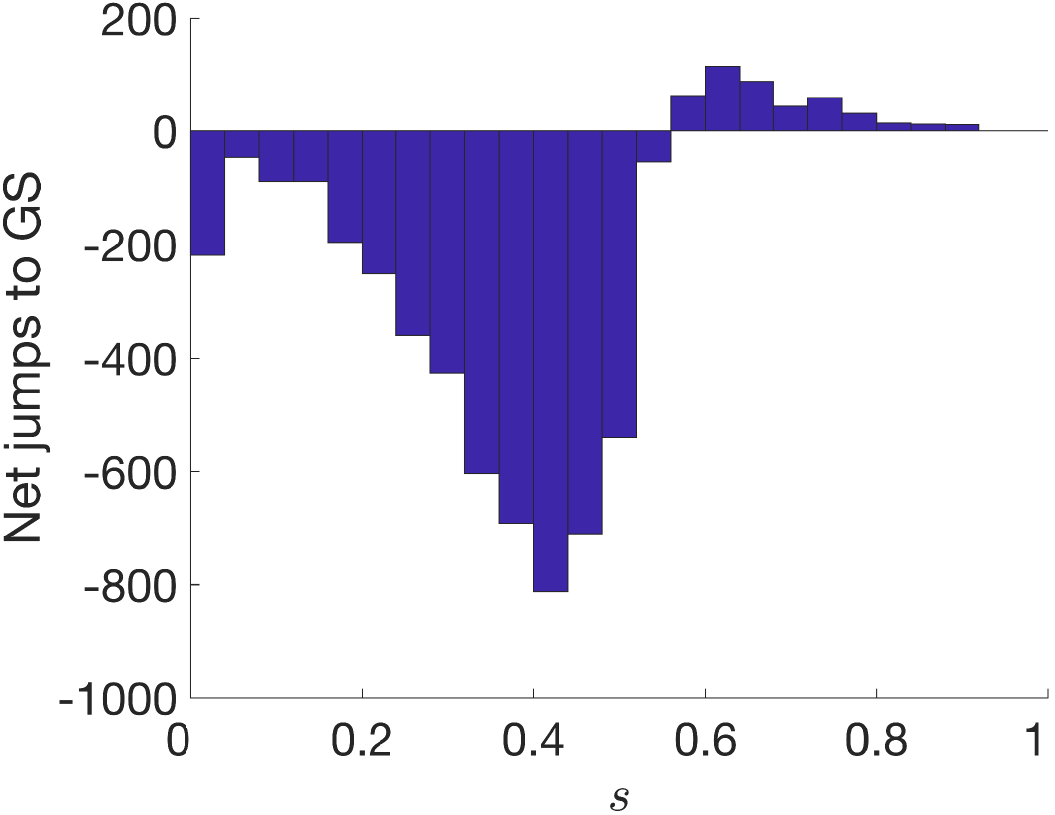}\label{fig:5c}} 
\caption{(a): Dotted lines: Linear schedule; Solid lines: D-Wave schedule used in Fig.~\ref{fig:3qubits8qubits}; (b): Same as in Fig.~\ref{fig:3qubits8qubits} (the evolution of the population in the instantaneous ground state for the $8$-qubit problem), but with a linear schedule. Inset: the convergence of the ground state population towards the AME results. (c): A histogram of the net number of jumps to the instantaneous ground state (GS). It shares the same pattern as in the non-linear schedule case (the inset of Fig.~\ref{fig:8qubitssingle}), but the net number of jumps out of the ground state is smaller due to the smaller energy scale in the linear schedule.}
\label{fig:8qubitslinear}
\end{figure*}

\section{Proof that eigenstates of $H_S$ are modified only under jumps}
\label{app:eigenstate-proof}

Recall that the Lindblad operators are defined in Eq.~\eqref{eq:Lindblad2} of the main text as:
\begin{align}
L_{\alpha, \omega}(t) &= \sum_{a,b} \delta_{\omega, \eps_b(t) - \eps_a(t)} \bra{\eps_a(t)} A_\alpha \ket{\eps_b(t)} 
| \eps_a(t) \rangle  \langle \eps_b(t)| \,.
\end{align}
After inverting Eq.~\eqref{eq:L} we have 
\begin{equation}
A_{i,\omega}(t) = \sum_{\alpha}u^{*}_{i,\alpha}(\omega)L_{\alpha, \omega}(t) 
\end{equation}
for the new Lindblad operators corresponding to a diagonalized $\gamma$ matrix of decay rates.

Assume for simplicity that the $\gamma$ matrix is already diagonal, so that $u$ is just an identity transformation and $i$ is a relabeling of $\alpha$; then
\begin{align}
A_{i,\omega}(t) &= \sum_{a,b} \delta_{\omega, \eps_b(t) - \eps_a(t)} \bra{\eps_a(t)} A_\alpha \ket{\eps_b(t)} 
| \eps_a(t) \rangle  \langle \eps_b(t)| \nonumber \\
A^{\dagger}_{i,\omega}(t) &= \sum_{a,b} \delta_{\omega, \eps_b(t) - \eps_a(t)} \bra{\eps_b(t)} A_\alpha \ket{\eps_a(t)} 
| \eps_b(t) \rangle  \langle \eps_a(t)| \,.
\end{align}

Consider the drift term 
\[
-\frac{1}{2}\sum_{i} \left[ A_{i}^{\dagger}(t)A_{i}(t) - \braket{A_{i}^{\dagger}(t)A_{i}(t)}\right]\ket{\psi(t)}dt\ .
\] 
Since $A_i(t)$ comes from the redefinition where the index $i$ includes the Bohr frequencies [Eq.~\eqref{eqt:Aired}]: $\sqrt{\gamma'_i(\omega)}A_{i,\omega}(t)\rightarrow A_i(t)$, the drift term becomes the following after we reintroduce the Bohr frequencies:
\onecolumngrid
\begin{equation}
- \frac{1}{2}\sum_{i}\sum_{\omega} \gamma'_i(\omega)\left[ A_{i,\omega}^{\dagger}(t)A_{i,\omega}(t) - \bra{\psi(t)} A_{i,\omega}^{\dagger}(t)A_{i,\omega}(t) \ket{\psi(t)} \right]\ket{\psi(t)}dt \,. 
\label{eq:explicit}
\end{equation}

\subsection{When $\ket{\psi(t)}$ is an eigenstate of $H_{S}(t)$}
If the $\ket{\psi(t)}$ is an eigenstate of $H_{S}(t)$ (denoted as $\ket{\eps_b(t)}$ here),
\begin{align}
A_{i,\omega}(t)\ket{\psi(t)} = A_{i,\omega}(t)\ket{\eps_b(t)} &=  \sum_{a} \delta_{\omega, \eps_b(t) - \eps_a(t)} \bra{\eps_a(t)} A_\alpha \ket{\eps_b(t)} 
| \eps_a(t) \rangle \nonumber \\
\bra{\psi(t)}A^{\dagger}_{i,\omega}(t) = \bra{\eps_b(t)}A^{\dagger}_{i,\omega}(t) &= \sum_{a} \delta_{\omega, \eps_b(t) - \eps_a(t)} \bra{\eps_b(t)} A_\alpha \ket{\eps_a(t)} \langle \eps_a(t)| \,.
\end{align}

Concentrate on the term of the parenthesis inside each summation of Eq.~\eqref{eq:explicit}, i.e.
\begin{equation}
\left[ A_{i,\omega}^{\dagger}(t)A_{i,\omega}(t) - \bra{\eps_b(t)} A_{i,\omega}^{\dagger}(t)A_{i,\omega}(t) \ket{\eps_b(t)}  \right]\ket{\eps_b(t)} \,.
\label{eq:paraterm}
\end{equation}
The first term is:
\begin{align}
&\phantom{{}={}} A_{i,\omega}^{\dagger}(t) A_{i,\omega}(t)\ket{\eps_b(t)}  \nonumber\\
&= \left(\sum_{a^{\prime},b^{\prime}} \delta_{\omega, \eps_{b^{\prime}}(t) - \eps_{a^{\prime}}(t)} \bra{\eps_{b^{\prime}}(t)} A_\alpha \ket{\eps_{a^{\prime}}(t)} 
| \eps_{b^{\prime}}(t) \rangle  \langle \eps_{a^{\prime}}(t)| \right) \sum_{a} \delta_{\omega, \eps_b(t) - \eps_a(t)} \bra{\eps_a(t)} A_\alpha \ket{\eps_b(t)} 
| \eps_a(t) \rangle \nonumber\\
&= \sum_{a,b^{\prime}} \delta_{\omega, \eps_{b^{\prime}}(t) - \eps_{a}(t)} \delta_{\omega, \eps_b(t) - \eps_a(t)}\bra{\eps_{b^{\prime}}(t)} A_\alpha \ket{\eps_{a}(t)}  \bra{\eps_a(t)} A_\alpha \ket{\eps_b(t)}
| \eps_{b^{\prime}}(t) \rangle  \nonumber\\
&= \sum_{a,b^{\prime}} \delta_{0, \eps_{b^{\prime}}(t) - \eps_{b}(t)} \delta_{\omega, \eps_b(t) - \eps_a(t)}\bra{\eps_{b^{\prime}}(t)} A_\alpha \ket{\eps_{a}(t)}  \bra{\eps_a(t)} A_\alpha \ket{\eps_b(t)}
| \eps_{b^{\prime}}(t) \rangle   \, ,
\label{eq:firstterm}
\end{align}
where the sum over $b^{\prime}$ denotes the sum over $\ket{\eps_{b^{\prime}}}$ sharing the same energy as $\ket{\eps_{b}}$.
The second term is:
\begin{align}
&\phantom{{}={}} \bra{\eps_b(t)} A_{i,\omega}^{\dagger}(t)A_{i,\omega}(t) \ket{\eps_b(t)} \ket{\eps_b(t)}  \nonumber\\
&= \left(\sum_{a^{\prime}} \delta_{\omega, \eps_b(t) - \eps_{a^{\prime}}(t)} \bra{\eps_b(t)} A_\alpha \ket{\eps_{a^{\prime}}(t)} \langle \eps_{a^{\prime}}(t)| \right) \left(\sum_{a} \delta_{\omega, \eps_b(t) - \eps_a(t)} \bra{\eps_a(t)} A_\alpha \ket{\eps_b(t)} 
| \eps_a(t) \rangle \right)\ket{\eps_b(t)} \nonumber\\
&= \left(\sum_{a} \delta_{\omega, \eps_b(t) - \eps_{a}(t)} \bra{\eps_b(t)} A_\alpha \ket{\eps_{a}(t)} \bra{\eps_a(t)} A_\alpha \ket{\eps_b(t)} \right)\ket{\eps_b(t)} \,.\nonumber\\
\label{eq:secondterm}
\end{align}

Subtracting Eq.~\eqref{eq:secondterm} from Eq.~\eqref{eq:firstterm} yields the drift term [Eq.~\eqref{eq:paraterm}], which is not zero, but a linear combination of degenerate eigenstates with the same energy $\eps_b(t)$. Before the jump happens the environment leads to the redistribution of $\ket{\eps_b(t)}$ to other states in the same energy manifold. (The Lamb shift $H_{\text{\text{LS}}}(t) = \sum_{i,\omega}S_{i}(\omega) A^\dagger_{i,\omega}(t) A_{i,\omega}(t) $ also yields the same effect.) Since they all share the same energy, this does not affect the overlap with the ground state. If the evolution by $H_{\text{S}}(t)$ is adiabatic, such a linear combination will stay in the same energy manifold and this explains the square-pulse like behavior in the overlapping with the ground state in Fig.~\ref{fig:8qubitssingle} of the main text.

\subsection{No degeneracy in $\eps_b(t)$}
If there are no degenerate states with energy $\eps_b(t)$, 
\begin{align}
&\phantom{{}={}} A_{i,\omega}^{\dagger}(t) A_{i,\omega}(t)\ket{\eps_b(t)}  = \sum_{a} \delta_{\omega, \eps_b(t) - \eps_a(t)}\bra{\eps_{b}(t)} A_\alpha \ket{\eps_{a}(t)}  \bra{\eps_a(t)} A_\alpha \ket{\eps_b(t)}
| \eps_{b}(t) \rangle 
\end{align}
This cancels with $\bra{\eps_b(t)} A_{i,\omega}^{\dagger}(t)A_{i,\omega}(t) \ket{\eps_b(t)} \ket{\eps_b(t)}$ [Eq.~\eqref{eq:secondterm}] and the drift term [Eq.~\eqref{eq:paraterm}] becomes zero.
\newline
\twocolumngrid

\section{Derivation of Eq.~\eqref{eq:inctunnrate}}
\label{app:lindbladfermi}

When the state is $\ket{\psi_1(t)}$, its jump rate $\lambda_{1 \rightarrow 0}(t) $ to $\ket{\psi_0(t)}$ comprises the summation of Lindblad terms responsible for the $1 \rightarrow 0$ transition:
\begin{equation}
\lambda_{1 \rightarrow 0}(t) = \sum_{\alpha \in \{1\rightarrow 0\}}\braket{{A}_\alpha^{\dagger}(t){A}_{\alpha}(t)} \,.
\end{equation}
The summation is over the number of qubits $n$. Since each qubit is coupled to its own environment with an independent noise source, the $\gamma$ matrix in Eq.~\eqref{eq:unitarytransform} is already diagonal. From Eq.~\eqref{eq:Lindblad2}
we know that
\begin{equation}
L_{\alpha, \omega_{10}}(t) = \sum_{a,b} \delta_{\omega_{10}, \eps_b(t) - \eps_a(t)} \bra{\eps_a(t)} \sigma^{z}_{\alpha} \ket{\eps_b(t)} | \eps_a(t) \rangle  \langle \eps_b(t)|    \,.
\end{equation}

Assume that Bohr frequency $\omega_{10}(t)$ is due only to the $1\rightarrow 0$ transition (even if it is not, the other terms would be annihilated by the matrix element $\langle \psi_1(t) |\dots |\psi_1(t) \rangle$). The Lindblad operators ${A}_{\alpha}(t)$ have the form:
\begin{equation}
{A}_{\alpha}(t) = \sqrt{\gamma_{\alpha}(\omega_{10})}\langle\psi_0(t)|\sigma^{z}_{\alpha} |\psi_1(t)\rangle| \psi_0(t) \rangle  \langle \psi_1(t)|  \,.
\end{equation}
Therefore:
\begin{align}
\lambda_{1 \rightarrow 0}(t) &= \sum_{\alpha}  \langle \psi_1(t) | {A}^{\dagger}_{\alpha}(t)  {A}_{\alpha}(t)   |\psi_1(t) \rangle    \\
&=  \sum_{\alpha=1}^{n} \gamma_\alpha(\omega_{10})|\langle\psi_0(t)|\sigma^{z}_{\alpha} |\psi_1(t)\rangle|^2 \nonumber  \,.
\end{align}
Here $\gamma_\alpha(\omega_{10})$ is evaluated with respect to the Ohmic spectral density.

\bibliography{refs}

\begin{thebibliography}{28}%
\makeatletter
\providecommand \@ifxundefined [1]{%
 \@ifx{#1\undefined}
}%
\providecommand \@ifnum [1]{%
 \ifnum #1\expandafter \@firstoftwo
 \else \expandafter \@secondoftwo
 \fi
}%
\providecommand \@ifx [1]{%
 \ifx #1\expandafter \@firstoftwo
 \else \expandafter \@secondoftwo
 \fi
}%
\providecommand \natexlab [1]{#1}%
\providecommand \enquote  [1]{``#1''}%
\providecommand \bibnamefont  [1]{#1}%
\providecommand \bibfnamefont [1]{#1}%
\providecommand \citenamefont [1]{#1}%
\providecommand \href@noop [0]{\@secondoftwo}%
\providecommand \href [0]{\begingroup \@sanitize@url \@href}%
\providecommand \@href[1]{\@@startlink{#1}\@@href}%
\providecommand \@@href[1]{\endgroup#1\@@endlink}%
\providecommand \@sanitize@url [0]{\catcode `\\12\catcode `\$12\catcode
  `\&12\catcode `\#12\catcode `\^12\catcode `\_12\catcode `\%12\relax}%
\providecommand \@@startlink[1]{}%
\providecommand \@@endlink[0]{}%
\providecommand \url  [0]{\begingroup\@sanitize@url \@url }%
\providecommand \@url [1]{\endgroup\@href {#1}{\urlprefix }}%
\providecommand \urlprefix  [0]{URL }%
\providecommand \Eprint [0]{\href }%
\providecommand \doibase [0]{http://dx.doi.org/}%
\providecommand \selectlanguage [0]{\@gobble}%
\providecommand \bibinfo  [0]{\@secondoftwo}%
\providecommand \bibfield  [0]{\@secondoftwo}%
\providecommand \translation [1]{[#1]}%
\providecommand \BibitemOpen [0]{}%
\providecommand \bibitemStop [0]{}%
\providecommand \bibitemNoStop [0]{.\EOS\space}%
\providecommand \EOS [0]{\spacefactor3000\relax}%
\providecommand \BibitemShut  [1]{\csname bibitem#1\endcsname}%
\let\auto@bib@innerbib\@empty
\bibitem [{\citenamefont {Wiseman}\ and\ \citenamefont
  {Milburn}(2010)}]{Wiseman:book}%
  \BibitemOpen
  \bibfield  {author} {\bibinfo {author} {\bibfnamefont {H.M.}\ \bibnamefont
  {Wiseman}}\ and\ \bibinfo {author} {\bibfnamefont {G.J.}\ \bibnamefont
  {Milburn}},\ }\href@noop {} {\emph {\bibinfo {title} {Quantum Measurement and
  Control}}}\ (\bibinfo  {publisher} {Cambridge University Press},\ \bibinfo
  {year} {2010})\BibitemShut {NoStop}%
\bibitem [{\citenamefont {Breuer}\ and\ \citenamefont
  {Petruccione}(2002)}]{Breuer:2002}%
  \BibitemOpen
  \bibfield  {author} {\bibinfo {author} {\bibfnamefont {Heinz-Peter}\
  \bibnamefont {Breuer}}\ and\ \bibinfo {author} {\bibfnamefont {Francesco}\
  \bibnamefont {Petruccione}},\ }\href@noop {} {\emph {\bibinfo {title} {The
  Theory of Open Quantum Systems}}}\ (\bibinfo  {publisher} {Oxford University
  Press},\ \bibinfo {year} {2002})\BibitemShut {NoStop}%
\bibitem [{\citenamefont {Carmichael}(2009)}]{carmichael2009statistical}%
  \BibitemOpen
  \bibfield  {author} {\bibinfo {author} {\bibfnamefont {Howard~J}\
  \bibnamefont {Carmichael}},\ }\href@noop {} {\emph {\bibinfo {title}
  {Statistical Methods in Quantum Optics 2: Non-Classical Fields}}}\ (\bibinfo
  {publisher} {Springer Science \& Business Media},\ \bibinfo {year}
  {2009})\BibitemShut {NoStop}%
\bibitem [{\citenamefont {Dum}\ \emph {et~al.}(1992)\citenamefont {Dum},
  \citenamefont {Parkins}, \citenamefont {Zoller},\ and\ \citenamefont
  {Gardiner}}]{dum1992monte}%
  \BibitemOpen
  \bibfield  {author} {\bibinfo {author} {\bibfnamefont {R}~\bibnamefont
  {Dum}}, \bibinfo {author} {\bibfnamefont {A.S.}\ \bibnamefont {Parkins}},
  \bibinfo {author} {\bibfnamefont {P}~\bibnamefont {Zoller}}, \ and\ \bibinfo
  {author} {\bibfnamefont {C.W.}\ \bibnamefont {Gardiner}},\ }\bibfield
  {title} {\enquote {\bibinfo {title} {{Monte Carlo simulation of master
  equations in quantum optics for vacuum, thermal, and squeezed reservoirs}},}\
  }\href {https://doi.org/10.1103%2Fphysreva.46.4382} {\bibfield  {journal}
  {\bibinfo  {journal} {{Phys. Rev. A}}\ }\textbf {\bibinfo {volume} {46}},\
  \bibinfo {pages} {4382} (\bibinfo {year} {1992})}\BibitemShut {NoStop}%
\bibitem [{\citenamefont {M{\o}lmer}\ \emph {et~al.}(1993)\citenamefont
  {M{\o}lmer}, \citenamefont {Castin},\ and\ \citenamefont
  {Dalibard}}]{molmer1993monte}%
  \BibitemOpen
  \bibfield  {author} {\bibinfo {author} {\bibfnamefont {Klaus}\ \bibnamefont
  {M{\o}lmer}}, \bibinfo {author} {\bibfnamefont {Yvan}\ \bibnamefont
  {Castin}}, \ and\ \bibinfo {author} {\bibfnamefont {Jean}\ \bibnamefont
  {Dalibard}},\ }\bibfield  {title} {\enquote {\bibinfo {title} {Monte carlo
  wave-function method in quantum optics},}\ }\href {\doibase
  10.1364/josab.10.000524} {\bibfield  {journal} {\bibinfo  {journal} {JOSA B}\
  }\textbf {\bibinfo {volume} {10}},\ \bibinfo {pages} {524--538} (\bibinfo
  {year} {1993})}\BibitemShut {NoStop}%
\bibitem [{\citenamefont {Daley}(2014)}]{daley2014quantum}%
  \BibitemOpen
  \bibfield  {author} {\bibinfo {author} {\bibfnamefont {Andrew~J}\
  \bibnamefont {Daley}},\ }\bibfield  {title} {\enquote {\bibinfo {title}
  {Quantum trajectories and open many-body quantum systems},}\ }\href {\doibase
  10.1080/00018732.2014.933502} {\bibfield  {journal} {\bibinfo  {journal}
  {Advances in Physics}\ }\textbf {\bibinfo {volume} {63}},\ \bibinfo {pages}
  {77--149} (\bibinfo {year} {2014})}\BibitemShut {NoStop}%
\bibitem [{\citenamefont {Brun}(2002)}]{brun:719}%
  \BibitemOpen
  \bibfield  {author} {\bibinfo {author} {\bibfnamefont {Todd~A.}\ \bibnamefont
  {Brun}},\ }\bibfield  {title} {\enquote {\bibinfo {title} {A simple model of
  quantum trajectories},}\ }\href {\doibase 10.1119/1.1475328} {\bibfield
  {journal} {\bibinfo  {journal} {Am. J. Phys.}\ }\textbf {\bibinfo {volume}
  {70}},\ \bibinfo {pages} {719--737} (\bibinfo {year} {2002})}\BibitemShut
  {NoStop}%
\bibitem [{\citenamefont {Imamoglu}(1994)}]{Imamoglu:94}%
  \BibitemOpen
  \bibfield  {author} {\bibinfo {author} {\bibfnamefont {A.}~\bibnamefont
  {Imamoglu}},\ }\bibfield  {title} {\enquote {\bibinfo {title} {{Stochastic
  wave-function approach to non-Markovian systems}},}\ }\href
  {https://link.aps.org/doi/10.1103/PhysRevA.50.3650} {\bibfield  {journal}
  {\bibinfo  {journal} {Physical Review A}\ }\textbf {\bibinfo {volume} {50}},\
  \bibinfo {pages} {3650--3653} (\bibinfo {year} {1994})}\BibitemShut {NoStop}%
\bibitem [{\citenamefont {Breuer}(2004)}]{Breuer:2004wq}%
  \BibitemOpen
  \bibfield  {author} {\bibinfo {author} {\bibfnamefont {Heinz-Peter}\
  \bibnamefont {Breuer}},\ }\bibfield  {title} {\enquote {\bibinfo {title}
  {Genuine quantum trajectories for non-markovian processes},}\ }\href
  {http://link.aps.org/doi/10.1103/PhysRevA.70.012106} {\bibfield  {journal}
  {\bibinfo  {journal} {{Phys. Rev. A}}\ }\textbf {\bibinfo {volume} {70}},\
  \bibinfo {pages} {012106--} (\bibinfo {year} {2004})}\BibitemShut {NoStop}%
\bibitem [{\citenamefont {Caiaffa}\ \emph {et~al.}(2017)\citenamefont
  {Caiaffa}, \citenamefont {Smirne},\ and\ \citenamefont
  {Bassi}}]{Caiaffa:2017aa}%
  \BibitemOpen
  \bibfield  {author} {\bibinfo {author} {\bibfnamefont {Matteo}\ \bibnamefont
  {Caiaffa}}, \bibinfo {author} {\bibfnamefont {Andrea}\ \bibnamefont
  {Smirne}}, \ and\ \bibinfo {author} {\bibfnamefont {Angelo}\ \bibnamefont
  {Bassi}},\ }\bibfield  {title} {\enquote {\bibinfo {title} {Stochastic
  unraveling of positive quantum dynamics},}\ }\href
  {https://link.aps.org/doi/10.1103/PhysRevA.95.062101} {\bibfield  {journal}
  {\bibinfo  {journal} {Physical Review A}\ }\textbf {\bibinfo {volume} {95}},\
  \bibinfo {pages} {062101--} (\bibinfo {year} {2017})}\BibitemShut {NoStop}%
\bibitem [{\citenamefont {Davies}\ and\ \citenamefont
  {Spohn}(1978)}]{springerlink:10.1007/BF01011696}%
  \BibitemOpen
  \bibfield  {author} {\bibinfo {author} {\bibfnamefont {E.~B.}\ \bibnamefont
  {Davies}}\ and\ \bibinfo {author} {\bibfnamefont {H.}~\bibnamefont {Spohn}},\
  }\bibfield  {title} {\enquote {\bibinfo {title} {Open quantum systems with
  time-dependent hamiltonians and their linear response},}\ }\href {\doibase
  dx.doi.org/10.1007/BF01011696} {\bibfield  {journal} {\bibinfo  {journal} {J.
  Stat. Phys.}\ }\textbf {\bibinfo {volume} {19}},\ \bibinfo {pages} {511--523}
  (\bibinfo {year} {1978})}\BibitemShut {NoStop}%
\bibitem [{\citenamefont {Albash}\ \emph {et~al.}(2012)\citenamefont {Albash},
  \citenamefont {Boixo}, \citenamefont {Lidar},\ and\ \citenamefont
  {Zanardi}}]{ABLZ:12-SI}%
  \BibitemOpen
  \bibfield  {author} {\bibinfo {author} {\bibfnamefont {Tameem}\ \bibnamefont
  {Albash}}, \bibinfo {author} {\bibfnamefont {Sergio}\ \bibnamefont {Boixo}},
  \bibinfo {author} {\bibfnamefont {Daniel~A}\ \bibnamefont {Lidar}}, \ and\
  \bibinfo {author} {\bibfnamefont {Paolo}\ \bibnamefont {Zanardi}},\
  }\bibfield  {title} {\enquote {\bibinfo {title} {{Quantum adiabatic Markovian
  master equations}},}\ }\href {\doibase 10.1088/1367-2630/14/12/123016}
  {\bibfield  {journal} {\bibinfo  {journal} {New J. of Phys.}\ }\textbf
  {\bibinfo {volume} {14}},\ \bibinfo {pages} {123016} (\bibinfo {year}
  {2012})}\BibitemShut {NoStop}%
\bibitem [{\citenamefont {Das}\ and\ \citenamefont
  {Chakrabarti}(2008)}]{RevModPhys.80.1061}%
  \BibitemOpen
  \bibfield  {author} {\bibinfo {author} {\bibfnamefont {Arnab}\ \bibnamefont
  {Das}}\ and\ \bibinfo {author} {\bibfnamefont {Bikas~K.}\ \bibnamefont
  {Chakrabarti}},\ }\bibfield  {title} {\enquote {\bibinfo {title}
  {\textit{Colloquium}: Quantum annealing and analog quantum computation},}\
  }\href {\doibase 10.1103/RevModPhys.80.1061} {\bibfield  {journal} {\bibinfo
  {journal} {Rev. Mod. Phys.}\ }\textbf {\bibinfo {volume} {80}},\ \bibinfo
  {pages} {1061--1081} (\bibinfo {year} {2008})}\BibitemShut {NoStop}%
\bibitem [{\citenamefont {Albash}\ and\ \citenamefont
  {Lidar}(2016)}]{Albash-Lidar:RMP}%
  \BibitemOpen
  \bibfield  {author} {\bibinfo {author} {\bibfnamefont {Tameem}\ \bibnamefont
  {Albash}}\ and\ \bibinfo {author} {\bibfnamefont {Daniel~A.}\ \bibnamefont
  {Lidar}},\ }\bibfield  {title} {\enquote {\bibinfo {title} {Adiabatic quantum
  computing},}\ }\href {http://arXiv.org/abs/1611.04471} {\bibfield  {journal}
  {\bibinfo  {journal} {arXiv:1611.04471}\ } (\bibinfo {year}
  {2016})}\BibitemShut {NoStop}%
\bibitem [{\citenamefont {Santoro}\ \emph {et~al.}(2002)\citenamefont
  {Santoro}, \citenamefont {Marto\v{n}\'{a}k}, \citenamefont {Tosatti},\ and\
  \citenamefont {Car}}]{Santoro}%
  \BibitemOpen
  \bibfield  {author} {\bibinfo {author} {\bibfnamefont {Giuseppe~E.}\
  \bibnamefont {Santoro}}, \bibinfo {author} {\bibfnamefont {Roman}\
  \bibnamefont {Marto\v{n}\'{a}k}}, \bibinfo {author} {\bibfnamefont {Erio}\
  \bibnamefont {Tosatti}}, \ and\ \bibinfo {author} {\bibfnamefont {Roberto}\
  \bibnamefont {Car}},\ }\bibfield  {title} {\enquote {\bibinfo {title} {Theory
  of quantum annealing of an {I}sing spin glass},}\ }\href
  {http://science.sciencemag.org/content/295/5564/2427} {\bibfield  {journal}
  {\bibinfo  {journal} {Science}\ }\textbf {\bibinfo {volume} {295}},\ \bibinfo
  {pages} {2427--2430} (\bibinfo {year} {2002})}\BibitemShut {NoStop}%
\bibitem [{\citenamefont {Heim}\ \emph {et~al.}(2015)\citenamefont {Heim},
  \citenamefont {R{\o}nnow}, \citenamefont {Isakov},\ and\ \citenamefont
  {Troyer}}]{Heim:2014jf}%
  \BibitemOpen
  \bibfield  {author} {\bibinfo {author} {\bibfnamefont {Bettina}\ \bibnamefont
  {Heim}}, \bibinfo {author} {\bibfnamefont {Troels~F.}\ \bibnamefont
  {R{\o}nnow}}, \bibinfo {author} {\bibfnamefont {Sergei~V.}\ \bibnamefont
  {Isakov}}, \ and\ \bibinfo {author} {\bibfnamefont {Matthias}\ \bibnamefont
  {Troyer}},\ }\bibfield  {title} {\enquote {\bibinfo {title} {{Quantum versus
  classical annealing of Ising spin glasses}},}\ }\href
  {http://science.sciencemag.org/content/348/6231/215} {\bibfield  {journal}
  {\bibinfo  {journal} {Science}\ }\textbf {\bibinfo {volume} {348}},\ \bibinfo
  {pages} {215--217} (\bibinfo {year} {2015})}\BibitemShut {NoStop}%
\bibitem [{\citenamefont {Albash}\ and\ \citenamefont
  {Lidar}(2017)}]{Albash:2017aa}%
  \BibitemOpen
  \bibfield  {author} {\bibinfo {author} {\bibfnamefont {Tameem}\ \bibnamefont
  {Albash}}\ and\ \bibinfo {author} {\bibfnamefont {Daniel~A.}\ \bibnamefont
  {Lidar}},\ }\bibfield  {title} {\enquote {\bibinfo {title} {Evidence for a
  limited quantum speedup on a quantum annealer},}\ }\href
  {http://arXiv.org/abs/1705.07452} {\bibfield  {journal} {\bibinfo  {journal}
  {arXiv:1705.07452}\ } (\bibinfo {year} {2017})}\BibitemShut {NoStop}%
\bibitem [{\citenamefont {Albash}\ and\ \citenamefont
  {Lidar}(2015)}]{Albash:2015nx}%
  \BibitemOpen
  \bibfield  {author} {\bibinfo {author} {\bibfnamefont {Tameem}\ \bibnamefont
  {Albash}}\ and\ \bibinfo {author} {\bibfnamefont {Daniel~A.}\ \bibnamefont
  {Lidar}},\ }\bibfield  {title} {\enquote {\bibinfo {title} {Decoherence in
  adiabatic quantum computation},}\ }\href
  {http://link.aps.org/doi/10.1103/PhysRevA.91.062320} {\bibfield  {journal}
  {\bibinfo  {journal} {Phys. Rev. A}\ }\textbf {\bibinfo {volume} {91}},\
  \bibinfo {pages} {062320--} (\bibinfo {year} {2015})}\BibitemShut {NoStop}%
\bibitem [{\citenamefont {Laine}\ \emph {et~al.}(2010)\citenamefont {Laine},
  \citenamefont {Piilo},\ and\ \citenamefont {Breuer}}]{laine2010measure}%
  \BibitemOpen
  \bibfield  {author} {\bibinfo {author} {\bibfnamefont {Elsi-Mari}\
  \bibnamefont {Laine}}, \bibinfo {author} {\bibfnamefont {Jyrki}\ \bibnamefont
  {Piilo}}, \ and\ \bibinfo {author} {\bibfnamefont {Heinz-Peter}\ \bibnamefont
  {Breuer}},\ }\bibfield  {title} {\enquote {\bibinfo {title} {{Measure for the
  non-Markovianity of quantum processes}},}\ }\href
  {https://journals.aps.org/pra/abstract/10.1103/PhysRevA.81.062115} {\bibfield
   {journal} {\bibinfo  {journal} {Physical Review A}\ }\textbf {\bibinfo
  {volume} {81}},\ \bibinfo {pages} {062115} (\bibinfo {year}
  {2010})}\BibitemShut {NoStop}%
\bibitem [{\citenamefont {Gardiner}\ and\ \citenamefont
  {Zoller}(2004)}]{gardiner2004quantum}%
  \BibitemOpen
  \bibfield  {author} {\bibinfo {author} {\bibfnamefont {C.}~\bibnamefont
  {Gardiner}}\ and\ \bibinfo {author} {\bibfnamefont {P.}~\bibnamefont
  {Zoller}},\ }\href {http://books.google.com/books?id=a\_xsT8oGhdgC} {\emph
  {\bibinfo {title} {Quantum Noise: A Handbook of Markovian and Non-Markovian
  Quantum Stochastic Methods with Applications to Quantum Optics}}},\ Springer
  Series in Synergetics\ (\bibinfo  {publisher} {Springer},\ \bibinfo {year}
  {2004})\BibitemShut {NoStop}%
\bibitem [{\citenamefont {Breuer}\ and\ \citenamefont
  {Piilo}(2009)}]{breuer2009stochastic}%
  \BibitemOpen
  \bibfield  {author} {\bibinfo {author} {\bibfnamefont {H-P}\ \bibnamefont
  {Breuer}}\ and\ \bibinfo {author} {\bibfnamefont {Jyrki}\ \bibnamefont
  {Piilo}},\ }\bibfield  {title} {\enquote {\bibinfo {title} {Stochastic jump
  processes for non-markovian quantum dynamics},}\ }\href
  {http://iopscience.iop.org/article/10.1209/0295-5075/85/50004/fulltext/}
  {\bibfield  {journal} {\bibinfo  {journal} {EPL (Europhysics Letters)}\
  }\textbf {\bibinfo {volume} {85}},\ \bibinfo {pages} {50004} (\bibinfo {year}
  {2009})}\BibitemShut {NoStop}%
\bibitem [{\citenamefont {Piilo}\ \emph {et~al.}(2008)\citenamefont {Piilo},
  \citenamefont {Maniscalco}, \citenamefont {H{\"a}rk{\"o}nen},\ and\
  \citenamefont {Suominen}}]{piilo2008non}%
  \BibitemOpen
  \bibfield  {author} {\bibinfo {author} {\bibfnamefont {Jyrki}\ \bibnamefont
  {Piilo}}, \bibinfo {author} {\bibfnamefont {Sabrina}\ \bibnamefont
  {Maniscalco}}, \bibinfo {author} {\bibfnamefont {Kari}\ \bibnamefont
  {H{\"a}rk{\"o}nen}}, \ and\ \bibinfo {author} {\bibfnamefont {Kalle-Antti}\
  \bibnamefont {Suominen}},\ }\bibfield  {title} {\enquote {\bibinfo {title}
  {Non-markovian quantum jumps},}\ }\href
  {https://link.aps.org/doi/10.1103/PhysRevLett.100.180402} {\bibfield
  {journal} {\bibinfo  {journal} {Physical Review Letters}\ }\textbf {\bibinfo
  {volume} {100}},\ \bibinfo {pages} {180402--} (\bibinfo {year}
  {2008})}\BibitemShut {NoStop}%
\bibitem [{\citenamefont {Piilo}\ \emph {et~al.}(2009)\citenamefont {Piilo},
  \citenamefont {H{\"a}rk{\"o}nen}, \citenamefont {Maniscalco},\ and\
  \citenamefont {Suominen}}]{piilo2009open}%
  \BibitemOpen
  \bibfield  {author} {\bibinfo {author} {\bibfnamefont {Jyrki}\ \bibnamefont
  {Piilo}}, \bibinfo {author} {\bibfnamefont {K}~\bibnamefont
  {H{\"a}rk{\"o}nen}}, \bibinfo {author} {\bibfnamefont {Sabrina}\ \bibnamefont
  {Maniscalco}}, \ and\ \bibinfo {author} {\bibfnamefont {K-A}\ \bibnamefont
  {Suominen}},\ }\bibfield  {title} {\enquote {\bibinfo {title} {Open system
  dynamics with non-markovian quantum jumps},}\ }\href
  {https://journals.aps.org/pra/abstract/10.1103/PhysRevA.79.062112} {\bibfield
   {journal} {\bibinfo  {journal} {Physical Review A}\ }\textbf {\bibinfo
  {volume} {79}},\ \bibinfo {pages} {062112} (\bibinfo {year}
  {2009})}\BibitemShut {NoStop}%
\bibitem [{\citenamefont {H{\"a}rk{\"o}nen}(2010)}]{harkonen2010jump}%
  \BibitemOpen
  \bibfield  {author} {\bibinfo {author} {\bibfnamefont {Kari}\ \bibnamefont
  {H{\"a}rk{\"o}nen}},\ }\bibfield  {title} {\enquote {\bibinfo {title} {Jump
  probabilities in the non-markovian quantum jump method},}\ }\href
  {http://iopscience.iop.org/article/10.1088/1751-8113/43/6/065302/meta}
  {\bibfield  {journal} {\bibinfo  {journal} {Journal of Physics A:
  Mathematical and Theoretical}\ }\textbf {\bibinfo {volume} {43}},\ \bibinfo
  {pages} {065302} (\bibinfo {year} {2010})}\BibitemShut {NoStop}%
\bibitem [{\citenamefont {Boixo}\ \emph {et~al.}(2016)\citenamefont {Boixo},
  \citenamefont {Smelyanskiy}, \citenamefont {Shabani}, \citenamefont {Isakov},
  \citenamefont {Dykman}, \citenamefont {Denchev}, \citenamefont {Amin},
  \citenamefont {Smirnov}, \citenamefont {Mohseni},\ and\ \citenamefont
  {Neven}}]{Boixo:2014yu}%
  \BibitemOpen
  \bibfield  {author} {\bibinfo {author} {\bibfnamefont {Sergio}\ \bibnamefont
  {Boixo}}, \bibinfo {author} {\bibfnamefont {Vadim~N.}\ \bibnamefont
  {Smelyanskiy}}, \bibinfo {author} {\bibfnamefont {Alireza}\ \bibnamefont
  {Shabani}}, \bibinfo {author} {\bibfnamefont {Sergei~V.}\ \bibnamefont
  {Isakov}}, \bibinfo {author} {\bibfnamefont {Mark}\ \bibnamefont {Dykman}},
  \bibinfo {author} {\bibfnamefont {Vasil~S.}\ \bibnamefont {Denchev}},
  \bibinfo {author} {\bibfnamefont {Mohammad~H.}\ \bibnamefont {Amin}},
  \bibinfo {author} {\bibfnamefont {Anatoly~Yu}\ \bibnamefont {Smirnov}},
  \bibinfo {author} {\bibfnamefont {Masoud}\ \bibnamefont {Mohseni}}, \ and\
  \bibinfo {author} {\bibfnamefont {Hartmut}\ \bibnamefont {Neven}},\
  }\bibfield  {title} {\enquote {\bibinfo {title} {Computational multiqubit
  tunnelling in programmable quantum annealers},}\ }\href
  {http://dx.doi.org/10.1038/ncomms10327} {\bibfield  {journal} {\bibinfo
  {journal} {Nat Commun}\ }\textbf {\bibinfo {volume} {7}} (\bibinfo {year}
  {2016})}\BibitemShut {NoStop}%
\bibitem [{\citenamefont {Boixo}\ \emph {et~al.}(2014)\citenamefont {Boixo},
  \citenamefont {Ronnow}, \citenamefont {Isakov}, \citenamefont {Wang},
  \citenamefont {Wecker}, \citenamefont {Lidar}, \citenamefont {Martinis},\
  and\ \citenamefont {Troyer}}]{q108}%
  \BibitemOpen
  \bibfield  {author} {\bibinfo {author} {\bibfnamefont {Sergio}\ \bibnamefont
  {Boixo}}, \bibinfo {author} {\bibfnamefont {Troels~F.}\ \bibnamefont
  {Ronnow}}, \bibinfo {author} {\bibfnamefont {Sergei~V.}\ \bibnamefont
  {Isakov}}, \bibinfo {author} {\bibfnamefont {Zhihui}\ \bibnamefont {Wang}},
  \bibinfo {author} {\bibfnamefont {David}\ \bibnamefont {Wecker}}, \bibinfo
  {author} {\bibfnamefont {Daniel~A.}\ \bibnamefont {Lidar}}, \bibinfo {author}
  {\bibfnamefont {John~M.}\ \bibnamefont {Martinis}}, \ and\ \bibinfo {author}
  {\bibfnamefont {Matthias}\ \bibnamefont {Troyer}},\ }\bibfield  {title}
  {\enquote {\bibinfo {title} {Evidence for quantum annealing with more than
  one hundred qubits},}\ }\href {\doibase 10.1038/nphys2900} {\bibfield
  {journal} {\bibinfo  {journal} {Nat. Phys.}\ }\textbf {\bibinfo {volume}
  {10}},\ \bibinfo {pages} {218--224} (\bibinfo {year} {2014})}\BibitemShut
  {NoStop}%
\bibitem [{\citenamefont {Gao}\ \emph {et~al.}(2015)\citenamefont {Gao},
  \citenamefont {Neuhauser}, \citenamefont {Baer},\ and\ \citenamefont
  {Rabani}}]{gao2015sublinear}%
  \BibitemOpen
  \bibfield  {author} {\bibinfo {author} {\bibfnamefont {Yi}~\bibnamefont
  {Gao}}, \bibinfo {author} {\bibfnamefont {Daniel}\ \bibnamefont {Neuhauser}},
  \bibinfo {author} {\bibfnamefont {Roi}\ \bibnamefont {Baer}}, \ and\ \bibinfo
  {author} {\bibfnamefont {Eran}\ \bibnamefont {Rabani}},\ }\bibfield  {title}
  {\enquote {\bibinfo {title} {Sublinear scaling for time-dependent stochastic
  density functional theory},}\ }\href
  {http://aip.scitation.org/doi/pdf/10.1063/1.4905568} {\bibfield  {journal}
  {\bibinfo  {journal} {The Journal of chemical physics}\ }\textbf {\bibinfo
  {volume} {142}},\ \bibinfo {pages} {034106} (\bibinfo {year}
  {2015})}\BibitemShut {NoStop}%
\bibitem [{\citenamefont {Gillespie}\ \emph {et~al.}(1977)\citenamefont
  {Gillespie} \emph {et~al.}}]{gillespie1977exact}%
  \BibitemOpen
  \bibfield  {author} {\bibinfo {author} {\bibfnamefont {Daniel~T}\
  \bibnamefont {Gillespie}} \emph {et~al.},\ }\bibfield  {title} {\enquote
  {\bibinfo {title} {Exact stochastic simulation of coupled chemical
  reactions},}\ }\href {\doibase 10.1021/j100540a008} {\bibfield  {journal}
  {\bibinfo  {journal} {J. phys. Chem}\ }\textbf {\bibinfo {volume} {81}},\
  \bibinfo {pages} {2340--2361} (\bibinfo {year} {1977})}\BibitemShut {NoStop}%
\end{thebibliography}%
\end{document}